\PassOptionsToPackage{numbers,sort&compress}{natbib}
\documentclass[aps,prd,10pt,a4paper,nofootinbib,groupedaddress]{revtex4-2}

\usepackage{mathtools} 
\usepackage{amssymb}   
\usepackage{bm}
\usepackage{physics}
\usepackage{cancel}
\usepackage{empheq}
\usepackage{cases}

\usepackage{graphicx}
\usepackage{xcolor}
\usepackage{subcaption}
\usepackage{float}
\usepackage{placeins}
\usepackage{array}
\usepackage{tabularx}
\usepackage{multirow}
\usepackage{booktabs}

\usepackage{tikz}
\usetikzlibrary{arrows,positioning}

\usepackage{comment}
\usepackage{blindtext}
\usepackage{ulem}
\usepackage[bottom]{footmisc}
\usepackage{lineno}

\usepackage{orcidlink}
\hypersetup{
    colorlinks=true,
    linkcolor=blue,
    citecolor=blue,
    urlcolor=blue
}

\begin{document}

\title{Beyond Recoherence: Non-Markovianity and Operational Information
Recovery in Two-Field Inflation}

\author{Amritendu Banerjee\,\orcidlink{0009-0009-2281-2900}}
\email{amritendu.b@phy.rkmvu.ac.in}

\author{Anirban Chatterjee\,\orcidlink{0000-0002-8904-7133}}
\email[Corresponding author: ]{anirban.chatterjee@gm.rkmvu.ac.in}
\email[Alternative email: ]{iitkanirbanc@gmail.com}

\author{Atreya\,\orcidlink{0009-0007-2183-7853}}
\email{atreya@gm.rkmvu.ac.in}

\author{Abhijit Bandyopadhyay\,\orcidlink{0009-0009-4113-3356}}
\email{abhijit.phy@gm.rkmvu.ac.in}

\affiliation{Department of Physics, School of Mathematical Sciences,
Ramakrishna Mission Vivekananda Educational and Research Institute,
Belur Math, Howrah 711202, West Bengal, India}

\date{\today}

 \begin{abstract}
{\centering\textbf{Abstract}\par}
\smallskip
\noindent

 We investigate the dynamical correspondence between cosmological recoherence and operational information recovery in the reduced dynamics of primordial curvature perturbations during two-field inflation. In particular, we examine its dependence on the physical wave number of the curvature mode relative to the Hubble scale at the epoch of curvature–entropic interaction. At quadratic order in cosmological perturbations, the coupled curvature–entropic dynamics admits an exact Gaussian-channel description of the reduced curvature evolution. Within this framework, we employ revivals of trace distance and displacement-encoded quantum Fisher information (QFI) as complementary witnesses of information recovery, alongside reduced-state purity and CP divisibility. Our results reveal qualitatively distinct behavior across the super-Hubble, near-Hubble, and sub-Hubble interaction regimes. For the benchmark background, curvature–entropic entanglement is most pronounced near Hubble crossing, whereas sub-Hubble modes can exhibit an almost complete transient suppression of displacement sensitivity despite retaining nearly pure reduced states. The diagnostics further exhibit distinct temporal signatures and dependencies on the entropic mass, the turn amplitude of the background field-space trajectory, and the phase-space orientation of the initial Gaussian displacement encoding. These findings reveal a scale-dependent separation between cosmological recoherence, operational information recovery, and quantum non-Markovianity, demonstrating that the accessibility of initially encoded information during inflation cannot be inferred from reduced-state purity alone.
 \end{abstract}

\maketitle

\section{Introduction}
\label{sec:introduction} 
The quantum origin of primordial scalar 
perturbations is a cornerstone
of inflationary cosmology. 
The comoving curvature perturbation, which characterizes 
fluctuations in the intrinsic spatial
curvature of comoving hypersurfaces, 
provides a natural description
of these perturbations.  
During inflation, its Fourier modes become strongly squeezed,
particularly after Hubble crossing
\cite{AlbrechtEtAl1994,PolarskiStarobinsky1996}.
Such unitary
squeezing, however, should be distinguished from decoherence arising
from interactions with additional degrees of freedom
\cite{MukhanovFeldmanBrandenberger1992,PolarskiStarobinsky1996,
BurgessEtAl2023}. 
Multifield inflation provides a natural setting  to
investigate this distinction, since a turn of the homogeneous
background trajectory in field space induces couplings between
the adiabatic (curvature) and entropic sectors already at quadratic
order
\cite{GordonWandsBassettMaartens2001,ShanderaAgarwalKamal2018,ColasGrainVennin2022},
with the two-field model considered in this work constituting
the simplest realization, comprising one adiabatic and one
entropic degree of freedom associated with perturbations
tangent and orthogonal to the background trajectory,
respectively.  
At quadratic order, distinct Fourier sectors evolve
independently, while within each sector the curvature
and entropic modes constitute a coupled bipartite quantum
system
\cite{ProkopecRigopoulos2007,ColasGrainVennin2022}.
Although the joint evolution is unitary, the
dynamically generated entanglement between the two modes
generally renders the reduced curvature state mixed, even
when the joint state is initially pure and uncorrelated.
Thus, cubic and higher-order interactions are not necessary
for this mechanism of reduced-state decoherence, although
they introduce additional interaction channels
\cite{BurgessEtAl2023}. 
Within a given Fourier sector, such entanglement-induced
decoherence need not be irreversible, as the reduced
curvature state may subsequently recover part or all of
its lost purity, a phenomenon referred to as
recoherence
\cite{ColasGrainVennin2023,KranasGrainVennin2025}.
Moreover, since purity is a basis-independent measure of
mixedness, its reduction through entanglement does not
necessarily imply the suppression of quantum coherences
in a physically relevant basis and is therefore insufficient,
by itself, to establish the emergence of classicality in
primordial perturbations
\cite{Zurek2003,CampoParentani2008}. \\

Since the statistics of primordial curvature perturbations
are directly related to cosmological observables, 
we identify
the curvature sector as the system of interest and the
entropic sector, whose independent observational signatures
may be suppressed, as an intrinsic quantum environment
\cite{ProkopecRigopoulos2007,Planck2018Inflation}.
A distinctive feature of this cosmological setting is that
a turn of the background trajectory localized in time can
induce curvature--entropic coupling on sub-Hubble,
near-Hubble, or super-Hubble scales, depending on the
comoving wave number of the perturbation mode.
The reduced dynamics of the curvature sector may therefore
depend not only on the coupling strength and the effective
entropic mass but also on the epoch of interaction relative
to Hubble crossing.
The quadratic Hamiltonian governing the perturbations
admits an exact phase-space treatment of the joint
dynamics in terms of linear symplectic transformations,
while the reduced curvature dynamics is described by a
Gaussian channel for initially uncorrelated sectors with
a fixed Gaussian entropic state
\cite{ColasGrainVennin2022,WeedbrookEtAl2012,
KranasGrainVennin2025}.\\

Recent studies have investigated the conditions for
cosmological recoherence and its dependence on the
properties of the entropic sector, the inflationary
background, and the initial quantum state
\cite{ColasGrainVennin2023,BrahmaEtAl2025,
CieloEtAl2025,CieloEtAl2026}.
Although recoherence, understood as the recovery of
reduced-state purity, reflects a reduction in
curvature--entropic entanglement when the joint state
is pure, it does not necessarily imply a recovery of
information encoded in the initial curvature state.
A central objective of this work is to investigate the
dynamical correspondence between these distinct notions
of recovery and its dependence on the physical wave number
of the curvature mode relative to the Hubble scale at
the epoch of curvature--entropic interaction.
To this end, we consider the revival of distinguishability
between reduced curvature states evolving from distinct
initial preparations, as quantified by their trace distance
in accordance with the Breuer--Laine--Piilo (BLP) criterion
\cite{BreuerLainePiilo2009}, and the revival of quantum
Fisher information (QFI) with respect to a parameter
encoded in the initial curvature state
\cite{LuWangSun2010,AbiusoEtAl2023,ScandiEtAl2025},
as complementary witnesses of information recovery.
These witnesses of information recovery can be understood
within the broader framework of quantum non-Markovianity,
which encompasses distinct characterizations of memory
effects in reduced quantum dynamics
\cite{BreuerEtAl2016,LuWangSun2010}.
Complementary to these preparation-dependent witnesses,
the Rivas--Huelga--Plenio (RHP) criterion provides a
state-independent characterization of quantum
non-Markovianity based on the violation of complete
positive (CP) divisibility of the reduced dynamical map
\cite{RivasHuelgaPlenio2010,BreuerEtAl2016}.
Information recovery in the reduced curvature sector
should, however, be distinguished from genuine causal
information backflow from the entropic sector.
Neither purity recovery, revivals of trace distance or QFI,
nor CP-indivisibility alone suffices to establish such
a causal transfer; its identification requires additional
information about the causal structure of the underlying
multitime quantum dynamics
\cite{PollockEtAl2018,BuscemiEtAl2025}.\\

 For each Fourier sector, we employ pairs of Gaussian
curvature states with identical covariance matrices and
opposite displacements along the canonical phase-space
directions $q$ or $p$. Revivals of the corresponding trace
distances, $D_q$ and $D_p$, provide preparation-dependent
witnesses of BLP non-Markovianity without optimization
over all initial state pairs.
To compare the trace-distance and QFI diagnostics on an
equal footing, we employ the same family of displaced
Gaussian states, with $F_Q^{(\delta_q)}$ and
$F_Q^{(\delta_p)}$ denoting the QFI with respect to the
displacement parameters along the $q$ and $p$ directions,
respectively, keeping the initial covariance matrix and
the reduced dynamical map parameter independent.
Rather than estimating cosmological parameters
\cite{RotondoNambu2017,PiotrakEtAl2026}, we use the QFI
to quantify the accessibility of initially encoded
information through its operational connection to the
quantum Cram\'er--Rao bound
\cite{BraunsteinCaves1994,Paris2009}.
For these displacement encodings, the QFI is directly
related to the Uhlmann fidelity between distinct
preparations \cite{BanchiBraunsteinPirandola2015},
whereas the trace distance provides a generally
inequivalent measure of their distinguishability.\\

We numerically investigate the evolution of the reduced
curvature purity, the trace distances $D_q$ and $D_p$,
and the displacement-encoded QFI, $F_Q^{(\delta_q)}$ and
$F_Q^{(\delta_p)}$, during inflation, together with the
CP-divisibility properties of the reduced dynamical map.
The analysis is carried out within a phenomenological
inflationary background with a Gaussian turn-rate profile,
with benchmark values assigned to the slow-roll parameter,
the entropic mass relative to the Hubble scale, and the
amplitude and width of the turn, as detailed in
Sec.~\ref{sec:numerical_methodology}.
We vary the
Fourier-scale parameter $\kappa_t$, defined as the ratio
of the physical wave number to the Hubble scale at the
epoch of maximum turning rate of the background trajectory,
to probe curvature--entropic interactions in the
super-Hubble, near-Hubble, and sub-Hubble regimes within
the same inflationary background.
In addition to examining   
the temporal evolution of these diagnostics,
we quantify, for each Fourier scale, 
the accumulated revivals
of the trace distances $D_q$ and $D_p$ 
and the displacement-encoded
QFI, $F_Q^{(\delta_q)}$ and $F_Q^{(\delta_p)}$, 
together with
the fractions of their respective maximum 
losses recovered
by the final epoch. 
These measures, along with the
fraction of recovered purity, 
provide a quantitative basis
for assessing the dynamical correspondence 
between recoherence
and information recovery across the super-Hubble, 
near-Hubble,
and sub-Hubble interaction regimes.\\

 Our results reveal a striking separation between reduced-state
purity and the accessibility of initially encoded information,
with qualitatively distinct signatures across the three
Fourier-scale regimes, an overview of which is presented in
Fig.~\ref{fig:intro_loss_recovery}.
While modes that are super-Hubble at the peak of the turn
largely preserve their initial purity, distinguishability,
and displacement sensitivity, those near Hubble crossing
exhibit the strongest generation of curvature--entropic
entanglement among the representative scales considered.
In contrast, sub-Hubble modes exhibit an almost complete
transient suppression of distinguishability and displacement
sensitivity despite a negligible loss of purity.
Thus, even a nearly pure curvature state can retain remarkably
little accessible information about its initial preparation,
and the subsequent recovery of this information need not
track purity recovery.
The inequivalence of these diagnostics is further manifested
in their temporal behavior, with the epochs of maximal
suppression exhibiting a distinct dependence on the Fourier
scale and the displacement direction, as illustrated in
Fig.~\ref{fig:intro_epochs_minima}. Moreover, the intervals
of CP-indivisibility need not coincide with those of
recoherence or information recovery. 
These findings identify the physical wave number relative
to the Hubble scale at the epoch of interaction as a key
parameter governing the relationship between entanglement
generation, information loss and recovery, and quantum
non-Markovianity during inflation.\\  

The remainder of the paper is organized as follows.
Sec.~\ref{sec:twofield} introduces the two-field inflationary background
and derives the coupled curvature--entropic dynamics at quadratic order.
Sec.~\ref{sec:gaussian} formulates the joint Gaussian evolution and the
exact reduced Gaussian channel for the curvature mode, including the
characterization of reduced purity and recoherence.
Sec.~\ref{sec:nonmarkov} introduces the RHP and BLP criteria used to
characterize CP-indivisibility and preparation distinguishability,
while Sec.~\ref{sec:qfi} develops the displacement-encoded QFI and its
relation to fidelity and trace distance.
The numerical setup and diagnostics are described in
Sec.~\ref{sec:numerical_methodology}, and the corresponding results are
presented in Sec.~\ref{sec:numerical_results}.
The conclusions and outlook are summarized in
Sec.~\ref{sec:conclusion}.

\section{Two-field inflationary dynamics}\label{sec:twofield}
\subsection{Background geometry and adiabatic--entropic decomposition} 
Consider two canonical scalar fields $\phi_i$ ($i=1,2$)
minimally coupled to gravity, with action
\begin{eqnarray}
S &=& \int \dd^4x\,\sqrt{-g}\bigg[
\frac{M_{\rm Pl}^2}{2}R
-\frac12\delta^{ij}g^{\mu\nu}
\partial_\mu\phi_i\partial_\nu\phi_j
-V(\phi_1,\phi_2)
\bigg],
\label{eq:twofield_action}
\end{eqnarray}
where $M_{\rm Pl}=(8\pi G)^{-1/2}$ and $\hbar=c=1$.
For a spatially flat FLRW background,
$\dd s^2=a^2(\eta)(-\dd\eta^2+\dd\mathbf{x}^2)$,
the homogeneous scalar fields $\bar\phi_i(\eta)$ satisfy
\begin{eqnarray}
\bar\phi_i''+2\mathcal H\bar\phi_i'
+a^2V_{,i}=0,
\label{eq:background_eom}
\end{eqnarray}
where $V_{,i}\equiv\partial V/\partial\phi_i$,
$\eta$ is conformal time, $\mathcal H=a'/a$,
and a prime denotes differentiation with respect to $\eta$.
The background fields trace a trajectory
$\bar{\bm\phi}(\eta)=(\bar\phi_1,\bar\phi_2)$
in the two-dimensional scalar-field space.
The unit tangent and normal vectors to this trajectory are
\begin{eqnarray}
\bm e_\sigma
=
\frac{(\bar\phi_1',\bar\phi_2')}{\bar\phi'},
\qquad
\bm e_s
=
\frac{(-\bar\phi_2',\bar\phi_1')}{\bar\phi'}.
\label{eq:adiabatic_basis}
\end{eqnarray}
where $\bar\phi'=\sqrt{\bar\phi_1'^2+\bar\phi_2'^2}$.
The turning of the trajectory is characterized by
by the turn rate $\theta'$, defined through
\begin{eqnarray}
\mathbf e_\sigma'
&=&
\theta'\mathbf e_s,
\qquad
\theta'
=
-\frac{a^2V_{,s}}{\bar\phi'},
\qquad
V_{,s}
\equiv
e_s^{\,i}V_{,i},
\label{eq:turn_rate}
\end{eqnarray}
where the potential derivatives are evaluated along the background
trajectory \cite{GordonWandsBassettMaartens2001}.
A straight background trajectory corresponds to
$\theta'=0$, whereas a nonzero turning rate induces
coupling between adiabatic and entropic perturbations
at quadratic order.  
For scalar perturbations about a spatially flat FLRW background, the metric
is written as
\begin{eqnarray}
\dd s^2
&=&
a^2(\eta)\bigg[
-(1+2A)\dd\eta^2
+2\partial_iB\,\dd x^i\dd\eta
+\left\{
(1-2\psi)\delta_{ij}
+2\partial_i\partial_jE
\right\}\dd x^i\dd x^j
\bigg],
\label{eq:scalar_metric_perturbations}
\end{eqnarray}
where $A$ and $B$ are the scalar lapse and shift perturbations, $\psi$
is the curvature perturbation, and $E$ parametrizes the scalar shear.
The lapse and shift are nondynamical and are eliminated using the
linearized Einstein constraint equations. The scalar-field perturbations
are conveniently expressed in terms of the gauge-invariant variables $
Q_i = \delta\phi_i
+
\frac{\bar\phi_i'}{\mathcal H}\psi $,
whose projections along the adiabatic and entropic directions define the comoving-curvature and entropy perturbations,
\begin{eqnarray}
\mathcal R
=
\frac{\mathcal H}{\bar\phi'}\,
\bm e_\sigma\cdot\bm Q\, , \quad
\qquad
\mathcal F
=
\bm e_s\cdot\bm Q .
\label{eq:curvature_entropy}
\end{eqnarray}
After imposing the constraint equations, the scalar sector is thus
described by the two physical dynamical degrees of freedom
$\mathcal R$ and $\mathcal F$
\cite{MukhanovFeldmanBrandenberger1992,GordonWandsBassettMaartens2001}.

\subsection{Quadratic action and canonical Hamiltonian} 
Within the conventions adopted here, the quadratic action for
the adiabatic and entropic perturbations of two canonical scalar
fields can be expressed as~\cite{ColasGrainVennin2022}
\begin{eqnarray}
S^{(2)}
=
\frac{1}{2}\int d\eta\,d^3x
\bigg\{
z^2\left[
(\mathcal R')^2-(\nabla\mathcal R)^2
\right] 
+a^2\left[
(\mathcal F')^2-(\nabla\mathcal F)^2
-a^2M_{\mathcal F}^2\mathcal F^2
\right]
+2\mathcal C\,\mathcal R'\mathcal F
\bigg\},
\label{eq:quadratic_action}
\end{eqnarray}
where $z \equiv \frac{a\bar\phi'}{\mathcal H}$
and the coefficients
$M_{\mathcal F}^2$ and $\mathcal C$ are given by
\begin{eqnarray}
M_{\mathcal F}^{\,2}
&=&
V_{,ss}-\frac{\theta'^{\,2}}{a^2},
\qquad
\mathcal C
=
2a^2\frac{\bar\phi'}{\mathcal H}\theta',
\label{eq:entropic_mass_coupling}
\end{eqnarray}
with $V_{,ss} \equiv e_s^{\,i}e_s^{\,j}
\left.
\frac{\partial^2 V}
{\partial\phi_i\partial\phi_j}
\right|_{\boldsymbol{\bar\phi}}$ 
denoting the projection of the potential Hessian along the
entropic direction.  
 The adiabatic--entropic coupling at quadratic order is governed by the turning rate $\theta'$ of the background trajectory in field space and vanishes for a straight trajectory ($\theta'=0$)~\cite{GordonWandsBassettMaartens2001,ColasGrainVennin2022}.
We introduce canonically normalized variables
$v_{\mathcal R}\equiv z\mathcal R$ and
$v_{\mathcal F}\equiv a\mathcal F$,
together with the strength of the derivative mixing   conveniently parametrized by
\begin{eqnarray}
\lambda(\eta)
&\equiv&
\frac{\mathcal C}{az}
=
2\theta'
=
2a\dot\theta,
\qquad
\alpha
\equiv
\frac{\dot\theta}{H}
=
\frac{\lambda}{2aH},
\label{eq:lambda_definition}
\end{eqnarray}
where $\lambda$ has dimensions of inverse conformal
time while $\alpha$ is a dimensionless measure of the turn rate.
The 
corresponding quadratic Hamiltonian
takes the form
\begin{eqnarray}
H^{(2)}
&=&
\frac{1}{2}\int d^3x
\Bigg[
\left(
\pi_{\mathcal R}-\lambda v_{\mathcal F}
\right)^2
+\pi_{\mathcal F}^{\,2}
+(\boldsymbol{\nabla}v_{\mathcal R})^2
+(\boldsymbol{\nabla}v_{\mathcal F})^2
-\frac{z''}{z}v_{\mathcal R}^2
\nonumber\\
&&
+\left(
a^2M_{\mathcal F}^{\,2}
-\frac{a''}{a}
\right)v_{\mathcal F}^2
+2\lambda\frac{z'}{z}
v_{\mathcal R}v_{\mathcal F}
\Bigg],
\label{eq:canonical_hamiltonian}
\end{eqnarray}
where $\pi_{\mathcal R} = v_{\mathcal R}'+\lambda v_{\mathcal F}$ and ,
$\pi_{\mathcal F} = v_{\mathcal F}'$
are the canonical momenta conjugate to
$v_{\mathcal R}$ and $v_{\mathcal F}$, respectively. \\

 Upon canonical quantization on the classical homogeneous background, the perturbation fields and their conjugate momenta are promoted to
  operators satisfying the equal-time commutation relations
$\left[\hat v_A(\eta,\mathbf x),\hat\pi_B(\eta,\mathbf y)\right]
=i\delta_{AB}\delta^{(3)}(\mathbf x-\mathbf y)$,
where $A,B\in\{\mathcal R,\mathcal F\}$.
Their Fourier components, $\hat v_{A,\mathbf k}(\eta)$ and $\hat\pi_{A,\mathbf k}(\eta)$, obey the Hermiticity conditions
$\hat v_{A,-\mathbf k}=\hat v_{A,\mathbf k}^{\dagger}$
and
$\hat\pi_{A,-\mathbf k}=\hat\pi_{A,\mathbf k}^{\dagger}$,
 implying that the $\mathbf k$ and $-\mathbf k$ modes are not independent degrees of freedom.  
For each nonzero momentum pair
$\{\mathbf k,-\mathbf k\}$, we introduce the
Hermitian cosine and sine quadratures
\begin{eqnarray}
\hat q_{A,k}^{(c)}
=
\frac{
\hat v_{A,\mathbf k}
+
\hat v_{A,-\mathbf k}
}{\sqrt{2}},
\quad
\hat p_{A,k}^{(c)}
=
\frac{
\hat \pi_{A,\mathbf k}
+
\hat \pi_{A,-\mathbf k}
}{\sqrt{2}},
\quad
\hat q_{A,k}^{(s)}
=
\frac{
\hat v_{A,\mathbf k}
-
\hat v_{A,-\mathbf k}
}{i\sqrt{2}},
\quad
\hat p_{A,k}^{(s)}
=
\frac{
\hat \pi_{A,\mathbf k}
-
\hat \pi_{A,-\mathbf k}
}{i\sqrt{2}} 
\label{eq:real_fourier_canonical_variables}
\end{eqnarray}
which satisfy 
$ [\hat q_{A,k}^{(r)},
\hat p_{B,k'}^{(r')}] 
= i\delta_{AB}\delta_{kk'}\delta_{rr'}$ 
with $r,r'\in\{c,s\}$.
Defining the time-dependent oscillator frequencies 
\begin{eqnarray}
\omega_{\mathcal R,k}^{2}
&\equiv&
k^2-\frac{z''}{z},
\qquad
\omega_{\mathcal F,k}^{2}
\equiv
k^2+a^2M_{\mathcal F}^{\,2}-\frac{a''}{a},
\label{eq:mode_frequencies_main}
\end{eqnarray}
the  quadratic Hamiltonian decomposes as
\begin{eqnarray}
H^{(2)}
&=&
\int_{\mathcal K_+}
\frac{d^3k}{(2\pi)^3}
\sum_{r=c,s}
H_k^{(r)},
\label{eq:real_sector_hamiltonian_decomposition}
\end{eqnarray}
where $\mathcal K_+$ contains one representative of each
$\{\mathbf k,-\mathbf k\}$ pair, and
\begin{eqnarray}
H_k^{(r)}
=
\frac{1}{2}
\Bigg[
\left(
p_{\mathcal R,k}^{(r)}
-
\lambda q_{\mathcal F,k}^{(r)}
\right)^2
+
\left(
p_{\mathcal F,k}^{(r)}
\right)^2
+
\omega_{\mathcal R,k}^{2}
\left(
q_{\mathcal R,k}^{(r)}
\right)^2 
+
\omega_{\mathcal F,k}^{2}
\left(
q_{\mathcal F,k}^{(r)}
\right)^2
+
2\lambda\frac{z'}{z}
q_{\mathcal R,k}^{(r)}
q_{\mathcal F,k}^{(r)}
\Bigg].
\label{eq:real_sector_hamiltonian}
\end{eqnarray}
Since the cosine and sine sectors evolve independently
under identical Hamiltonians, we consider a single real
sector for each momentum pair and suppress its label,
where the canonical
quadratures satisfy $ [\hat q_{A,k},
\hat p_{B,k'}] = i\delta_{AB}\delta_{kk'}$. 
The corresponding mode Hamiltonian decomposes as 
\begin{eqnarray}
\hat H_k
&=&
\hat H_{\mathcal R,k}
+
\hat H_{\mathcal F,k}
+
\hat H_{{\rm int},k},
\label{eq:mode_hamiltonian_decomposition}
\end{eqnarray}
with
\begin{eqnarray}
\hat H_{\mathcal R,k}
&=&
\frac{1}{2}
\left[
\hat p_{\mathcal R,k}^{\,2}
+
\omega_{\mathcal R,k}^{\,2}
\hat q_{\mathcal R,k}^{\,2}
\right],
\nonumber\\
\hat H_{\mathcal F,k}
&=&
\frac{1}{2}
\left[
\hat p_{\mathcal F,k}^{\,2}
+
\left(
\omega_{\mathcal F,k}^{\,2}
+\lambda^2
\right)
\hat q_{\mathcal F,k}^{\,2}
\right],
\nonumber\\
\hat H_{{\rm int},k}
&=&
-\lambda
\hat p_{\mathcal R,k}
\hat q_{\mathcal F,k}
+
\lambda\frac{z'}{z}
\hat q_{\mathcal R,k}
\hat q_{\mathcal F,k}.
\label{eq:mode_hamiltonian_components}
\end{eqnarray}
For a given representative real Fourier mode $k$, the quadratic theory
reduces to two linearly coupled canonical oscillators. 
The corresponding curvature
mode is identified as the system $\mathcal S_k$, while the entropic mode
is treated as the environment $\mathcal E_k$. 
 The density operator $\hat\rho_k(\eta)$ of the joint
adiabatic--entropic state evolves unitarily according to
\begin{eqnarray}
i\frac{d\hat\rho_k}{d\eta}
&=&
[\hat H_k,\hat\rho_k],
\label{eq:joint_density_evolution}
\end{eqnarray}
while the reduced curvature state is obtained by tracing out the
entropic mode,
\begin{eqnarray}
\hat\rho_{\mathcal S,k}(\eta)
&\equiv&
{\rm Tr}_{\mathcal E}\,
\hat\rho_k(\eta).
\label{eq:reduced_density_operator}
\end{eqnarray}
Tracing out the entropic mode renders 
the reduced curvature dynamics generally
nonunitary, as the interaction generates 
curvature--entropic
correlations
\cite{Boyanovsky2015,ShanderaAgarwalKamal2018,ColasGrainVennin2022}.

\section{Gaussian states and exact reduced curvature channel}\label{sec:gaussian}
\subsection{Symplectic evolution and state moments}
Collecting the canonical quadratures 
of the curvature and entropic
sectors into the phase-space vector
$\hat{\mathbf R}_k
 \equiv 
\left(
\hat q_{\mathcal R,k},
\hat p_{\mathcal R,k},
\hat q_{\mathcal F,k},
\hat p_{\mathcal F,k}
\right)^T$,
their canonical commutation 
relations take the symplectic form
\begin{eqnarray}
\left[
\hat R_{k,i},
\hat R_{k,j}
\right]
&=&
i \Omega_{ij},
\qquad
\Omega
=
\begin{pmatrix}
 J & 0\\
0 & J
\end{pmatrix},
\qquad
  J
=
\begin{pmatrix}
0 & 1\\
-1 & 0
\end{pmatrix}.
\label{eq:omega}
\end{eqnarray}
The mode Hamiltonian can then be written compactly as
$
\hat H_k(\eta)
=
\frac12
\hat{\mathbf R}_k^{\,T}
G_k(\eta)
\hat{\mathbf R}_k$,
where $G_k(\eta)$ is the real symmetric quadratic generator,
\begin{eqnarray}
G_k
&=&
\begin{pmatrix}
\omega_{\mathcal R,k}^{\,2} & 0
& \lambda z'/z & 0
\\
0 & 1 & -\lambda & 0
\\
\lambda z'/z & -\lambda
& \omega_{\mathcal F,k}^{\,2}+\lambda^2 & 0
\\
0 & 0 & 0 & 1
\end{pmatrix}.
\label{eq:G_matrix}
\end{eqnarray}
This representation makes explicit the symplectic structure of the
coupled curvature--entropic system and provides the natural starting
point for its Gaussian phase-space evolution. The
Heisenberg evolution follows from 
 $\hat{\bm R}'_k=  \Omega    G_k\hat{\bm R}_k$,
 whose solution
may be expressed as  
\begin{eqnarray}
\hat{\mathbf R}_k(\eta)
=
 M_k(\eta,\eta_0)\,
\hat{\mathbf R}_k(\eta_0),
\quad \mbox{with}\quad
  M_k'(\eta,\eta_0)
=   \Omega G_k(\eta)\,
 M_k(\eta,\eta_0),\quad
  M_k(\eta_0,\eta_0)
=
\mathbb{I}_4
\label{eq:phase_space_propagator}
\end{eqnarray}
 Preservation of the canonical commutation relations 
\eqref{eq:omega} 
requires $ M_k \Omega M_k^T =  \Omega $,
implying  $\det M_k=1$. 
The joint state $\rho_k$  involving
two canonical modes admits an equivalent representation in the four-dimensional
phase space spanned by
real phase space coordinates $
\mathbf R_k = \left(
q_{\mathcal R,k},
p_{\mathcal R,k},
q_{\mathcal F,k},
p_{\mathcal F,k}
\right)^T$
through its Wigner function $W(\mathbf R_k)$. 
 For a Gaussian state $\hat{\rho}_k$, the Wigner function takes the form
\begin{eqnarray}
W_k(\mathbf R_k)
&=&
\frac{
\exp\!\left[
-\frac{1}{2}
(\mathbf R_k-\mathbf d_k)^T
\boldsymbol{\sigma}_k^{-1}
(\mathbf R_k-\mathbf d_k)
\right]
}{
(2\pi)^2\sqrt{\det\boldsymbol{\sigma}_k}
},
\label{eq:gaussian_wigner_function}
\end{eqnarray}
where $\mathbf d_k$ 
is the displacement 
vector of first moments  and
$\boldsymbol{\sigma}_k$ is the 
covariance matrix of symmetrized 
second moments, defined by
\begin{eqnarray}
\mathbf d_k
&\equiv&
\left\langle \hat{\mathbf R}_k \right\rangle,
\qquad
(\boldsymbol{\sigma}_k)_{ij}
\equiv
\frac{1}{2}
\left\langle
\left\{
\Delta\hat R_{k,i},
\Delta\hat R_{k,j}
\right\}
\right\rangle ,
\label{eq:gaussian_moments}
\end{eqnarray}
where
$\Delta\hat R_{k,i}\equiv
\hat R_{k,i}-\langle\hat R_{k,i}\rangle$
and
$\langle \hat O\rangle\equiv\mathrm{Tr}(\hat{\rho}_k\hat O)$
denotes the expectation value of   with respect to $\hat{\rho}_k$. 
The covariance matrix of a physical Gaussian state satisfies the
Robertson--Schr\"odinger uncertainty relation
$\boldsymbol{\sigma}_k
+\frac{i}{2} \Omega 
 \geqslant 
0 $ which follows from the canonical commutation relations
in Eq.~(\ref{eq:omega}) and constitutes the physicality condition
for the Gaussian covariance matrix
\cite{WeedbrookEtAl2012,AdessoRagyLee2014}. 
For an initial Gaussian state $\hat{\rho}_k(\eta_0)$, the quadratic Hamiltonian
$\hat H_k$ preserves Gaussianity, and the joint dynamics is fully determined
by the evolution of the first and symmetrized second moments,
$\mathbf d_k(\eta)$ and $\boldsymbol{\sigma}_k(\eta)$
\cite{WeedbrookEtAl2012,AdessoRagyLee2014},
\begin{eqnarray}
\mathbf d_k(\eta)
=
  M_k(\eta,\eta_0)\,
\mathbf d_k(\eta_0),\qquad
\boldsymbol{\sigma}_k(\eta)
=
  M_k(\eta,\eta_0)\,
\boldsymbol{\sigma}_k(\eta_0)\,
  M_k^T(\eta,\eta_0).
\label{eq:exact_gaussian_evolution}
\end{eqnarray}

\subsection{Reduced Gaussian channel and complete positivity} 
To define the reduced dynamics of the curvature mode, we assume an
initially factorized state,
$
\hat\rho_k(\eta_0)
=
\hat\rho_{\mathcal S,k}(\eta_0)
\otimes
\hat\rho_{\mathcal E,k}(\eta_0)$,
with a fixed Gaussian preparation of the entropic sector. With respect
to the system--environment partition, the first moments, covariance
matrix, and symplectic propagator are decomposed as
\begin{eqnarray}
\mathbf d_k
=
\begin{pmatrix}
\mathbf d_{\mathcal S,k}\\
\mathbf d_{\mathcal E,k}
\end{pmatrix},
\quad
\boldsymbol{\sigma}_k
=
\begin{pmatrix}
\boldsymbol{\sigma}_{\mathcal{S},k}
&
\boldsymbol{\sigma}_{\mathcal{SE},k}
\\
\boldsymbol{\sigma}_{\mathcal{ES},k}
&
\boldsymbol{\sigma}_{\mathcal E,k}
\end{pmatrix},
\quad
M_k
=
\begin{pmatrix}
M_{\mathcal{S},k}
&
M_{\mathcal{SE},k}
\\
M_{\mathcal{ES},k}
&
M_{\mathcal E,k}
\end{pmatrix},
\end{eqnarray}
with the initial factorization implying
$\boldsymbol{\sigma}_k(\eta_0) = {\rm diag}(\boldsymbol{\sigma}_{\mathcal S,k}(\eta_0), \boldsymbol{\sigma}_{\mathcal E,k}(\eta_0))$. 
Using Eq.~(\ref{eq:exact_gaussian_evolution}), the reduced curvature
moments evolve as
\begin{eqnarray}
\mathbf d_{\mathcal S,k}(\eta)
&=&
X_k(\eta,\eta_0)\,
\mathbf d_{\mathcal S,k}(\eta_0)
+
c_k(\eta,\eta_0),
\label{eq:reduced_first_moments}
\\
\boldsymbol{\sigma}_{\mathcal S,k}(\eta)
&=&
X_k(\eta,\eta_0)\,
\boldsymbol{\sigma}_{\mathcal S,k}(\eta_0)\,
X_k^T(\eta,\eta_0)
+
Y_k(\eta,\eta_0),
\label{eq:reduced_covariance_channel}
\end{eqnarray}
where $X_k \equiv M_{\mathcal{S},k}$,
$Y_k \equiv 
M_{\mathcal{SE},k} 
\boldsymbol{\sigma}_{\mathcal E,k}(\eta_0) 
M_{\mathcal{SE},k}^{T}$ and $c_k
\equiv
M_{\mathcal{SE},k}\,
\mathbf d_{\mathcal E,k}(\eta_0) $  
define the exact adiabatic-mode Gaussian channel 
$\Phi_k(\eta,\eta_0)$ generated by 
tracing over the entropic mode,
\begin{eqnarray}
\hat\rho_{\mathcal S,k}(\eta)
&=&
{\rm Tr}_{\mathcal E}\,
\hat\rho_k(\eta) = \Phi_k(\eta,\eta_0)
\left[
\hat\rho_{\mathcal S,k}(\eta_0)
\right].
\label{eq:exact_reduced_map}
\end{eqnarray}
We assume a centered initial environmental state,
 $\mathbf d_{\mathcal E,k}(\eta_0)=0$ so that hence
 $\mathbf c_k=0$.  
The Wigner function of the reduced curvature state
$\hat\rho_{\mathcal S,k}$ is obtained by marginalizing over the
entropic phase space \cite{WeedbrookEtAl2012,AdessoRagyLee2014},
\begin{eqnarray}
W_{\mathcal S,k}(\mathbf R_{\mathcal S,k},\eta)
&=&
\int \dd^2\mathbf R_{\mathcal E,k}\,
W_k(\mathbf R_{\mathcal S,k},
\mathbf R_{\mathcal E,k};\eta)
\nonumber\\
&=&
\frac{
\exp\!\left[
-\frac12
\left(\mathbf R_{\mathcal S,k}-\mathbf d_{\mathcal S,k}\right)^T
\boldsymbol{\sigma}_{\mathcal S,k}^{-1}
\left(\mathbf R_{\mathcal S,k}-\mathbf d_{\mathcal S,k}\right)
\right]
}{
2\pi\sqrt{\det\boldsymbol{\sigma}_{\mathcal S,k}}
},
\label{eq:reduced_wigner_function}
\end{eqnarray}
where
$\mathbf R_{\mathcal S,k}=(q_{\mathcal R,k},p_{\mathcal R,k})^T$
and
$\mathbf R_{\mathcal E,k}=(q_{\mathcal F,k},p_{\mathcal F,k})^T$.  
The reduced map $\Phi_k(\eta,\eta_0)$ 
defines a physical Gaussian
channel when it is completely 
positive and trace preserving (CPTP).
For the covariance transformation in
Eq.~(\ref{eq:reduced_covariance_channel}), complete positivity requires
\cite{HolevoWerner2001,WeedbrookEtAl2012,AdessoRagyLee2014}
\begin{eqnarray}
Y_k
+\frac{i}{2}
\left(
J-X_k J X_k^T
\right)
&\geqslant&
0,
\qquad
J
=
\begin{pmatrix}
0&1\\
-1&0
\end{pmatrix},
\label{eq:gaussian_channel_CP}
\end{eqnarray}
which ensures preservation of the 
Robertson--Schr\"odinger uncertainty
relation $\boldsymbol{\sigma}_{\mathcal S, k}
+\frac{i}{2} J
 \geqslant 
0 $, for the reduced covariance matrix. 

\subsection{Decoherence from quadratic coupling  and recoherence}
\label{subsec:quadratic_decoherence}    
The bilinear couplings
$-\lambda\hat p_{\mathcal R,k}\hat q_{\mathcal F,k}$
and
$\lambda(z'/z)\hat q_{\mathcal R,k}\hat q_{\mathcal F,k}$
in Eq.~(\ref{eq:mode_hamiltonian_components})
can generate curvature--entropic correlations already
at quadratic order. Their effect on the reduced curvature
state can be characterized by its purity.
For Gaussian states, the joint and reduced purities
are respectively given by
\cite{WeedbrookEtAl2012,AdessoRagyLee2014}
\begin{eqnarray}
\mu_k(\eta)
\equiv 
{\rm Tr}\!\left[\hat\rho_k^2(\eta)\right]
=
\frac{1}{4\sqrt{\det\boldsymbol{\sigma}_k(\eta)}}\,\,,
\quad
\mu_{\mathcal S,k}(\eta)
 \equiv 
{\rm Tr}\!\left[\hat\rho_{\mathcal S,k}^2(\eta)\right]
=
\frac{1}{2\sqrt{\det\boldsymbol{\sigma}_{\mathcal S,k}(\eta)}}.
\label{eq:gaussian_purity}
\end{eqnarray}
Since $\det M_k=1$, the symplectic evolution in
Eq.~(\ref{eq:exact_gaussian_evolution}) preserves
$\det\boldsymbol{\sigma}_k$ and hence the joint
purity $\mu_k$. By contrast, the reduced covariance
evolves according to
Eq.~(\ref{eq:reduced_covariance_channel}), where
$X_k$ is generally nonsymplectic and $Y_k$ accounts
for the contribution of the entropic fluctuations.
Consequently, the reduced covariance determinant
and the corresponding purity $\mu_{\mathcal S,k}$
need not be conserved.\\

For an initially factorized pure joint state, a reduction of the curvature-mode purity to $\mu_{\mathcal S,k} \leqslant 1$ arises from entanglement generated with the entropic sector and, may be taken as a measure of reduced-state decoherence \cite{ColasGrainVennin2022,ColasGrainVennin2023,WeedbrookEtAl2012}. Here $\mu_{\mathcal S,k}$ refers to a single independent Fourier sector; since different $k$ sectors evolve independently at quadratic order, the resulting purity loss characterizes mode-resolved decoherence of the curvature perturbation.
Since purity is basis independent, this criterion does not by itself identify a preferred pointer basis or establish irreversible classicalization, namely the persistent emergence of effectively classical behavior through sustained suppression of quantum coherences and environment-induced state selection \cite{Zurek2003,Schlosshauer2005}. For a globally pure state, a subsequent increase in $\mu_{\mathcal S,k}$ corresponds to a reduction of curvature--entropic entanglement and is referred to here as recoherence \cite{ColasGrainVennin2023}. Such a purity revival, however, need not imply restoration of coherence in a specified basis or recovery of operationally accessible quantum information.  If the initially factorized curvature and entropic states are mixed, the reduced purity quantifies the overall mixedness of the curvature sector, including contributions from its initial mixedness and from subsequently generated intersector correlations, and therefore cannot by itself be used as a measure of only curvature--entropic entanglement.

\section{Operational criteria for quantum non-Markovianity}\label{sec:nonmarkov} 
Purity dynamics alone does not establish quantum non-Markovianity, which concerns memory effects in the reduced evolution and may manifest through the breakdown of CP-divisibility of the reduced dynamical map or through information backflow, identified by a temporary increase in the trace distance for a suitably chosen pair of reduced states evolving under the same dynamical map  \cite{RivasHuelgaPlenio2010,BreuerLainePiilo2009}. We therefore characterize the dynamics using complementary operational diagnostics based on CP-divisibility, revivals of state distinguishability, and the flow of quantum Fisher information, which probes the recovery of metrologically accessible information \cite{LuWangSun2010}.

\subsection{RHP: channel CP divisibility}
The Rivas--Huelga--Plenio (RHP) criterion characterizes quantum non-Markovianity through the
CP-divisibility of the reduced dynamical map
\cite{RivasHuelgaPlenio2010,BreuerEtAl2016}. Specifically, the family
$\Phi_k(\eta,\eta_0)$ is CP divisible if, for every
$\eta_2\geqslant\eta_1\geqslant\eta_0$, there exists a CPTP intermediate map
$\Phi_k(\eta_2,\eta_1)$ such that
\begin{eqnarray}
\Phi_k(\eta_2,\eta_0)
&=&
\Phi_k(\eta_2,\eta_1)
\circ
\Phi_k(\eta_1,\eta_0).
\label{eq:divisibility}
\end{eqnarray}
For invertible
$X_1\equiv X_k(\eta_1,\eta_0)$, the corresponding Gaussian
intermediate map is specified by
\begin{eqnarray}
X_{21}
&=&
X_2X_1^{-1},
\qquad
Y_{21}
=
Y_2-X_{21}Y_1X_{21}^{T},
\label{eq:intermediate}
\end{eqnarray}
where $X_2\equiv X_k(\eta_2,\eta_0)$ and
$Y_{1,2}\equiv Y_k(\eta_{1,2},\eta_0)$. Its complete positivity is
equivalent to the Gaussian-channel condition
\cite{TorreRogaIlluminati2015,HolevoWerner2001,WeedbrookEtAl2012}
\begin{eqnarray}
C_{21}
&\equiv&
Y_{21}
+\frac{i}{2}
\left(
J-X_{21}JX_{21}^{T}
\right)
\geqslant 0 .
\label{eq:intermediate_cp}
\end{eqnarray}
Hence, a negative eigenvalue of $C_{21}$ establishes CP-indivisibility
over the interval $(\eta_1,\eta_2)$
\cite{TorreRogaIlluminati2015,TorreIlluminati2018}. When $X_1$ is
singular, the inverse construction is not applicable and CP-divisibility
must instead be assessed directly from the existence of an admissible
intermediate Gaussian map. Numerically singular or ill-conditioned
points are therefore excluded from the inverse-based classification
rather than being identified as non-Markovian.\\

For numerical implementation, it is convenient to express the same
criterion locally in e-fold time $N\equiv\ln a$, with
$\dd/\dd N=(aH)^{-1}\dd/\dd\eta$. On intervals where $X(N)$ is
differentiable and invertible, define
\begin{eqnarray}
K(N)
&\equiv&
\left(\frac{\dd X}{\dd N}\right)X^{-1}.
\end{eqnarray}
For an infinitesimal interval $N\rightarrow N+\dd N$, the intermediate
Gaussian map becomes
\begin{eqnarray}
X_{N+\dd N,N}
&=&
\mathbf I
+
K\,\dd N
+
\mathcal O(\dd N^2),
\nonumber\\
Y_{N+\dd N,N}
&=&
\left(
\frac{\dd Y}{\dd N}
-KY-YK^T
\right)\dd N
+
\mathcal O(\dd N^2).
\end{eqnarray}
Substitution into Eq.~(\ref{eq:intermediate_cp}) yields, to leading
order,
\begin{eqnarray}
C_{\rm rate}
&\equiv&
\frac{\dd Y}{\dd N}
-KY-YK^T
-\frac{i}{2}
\left(
KJ+JK^T
\right)
\geqslant 0 .
\label{eq:cp_rate}
\end{eqnarray}
A negative eigenvalue of $C_{\rm rate}$ therefore provides a local
witness of CP-indivisibility, corresponding to the infinitesimal
Gaussian form of the RHP criterion
\cite{RivasHuelgaPlenio2010,TorreRogaIlluminati2015,
TorreIlluminati2018}.
Under a symplectic redefinition of the curvature quadratures,
$\mathbf R_{\mathcal S,k}\rightarrow
S\mathbf R_{\mathcal S,k}$, with $SJS^T=J$,
$C_{\rm rate}$ transforms by congruence,
$C_{\rm rate}\rightarrow S C_{\rm rate}S^T$
\cite{WeedbrookEtAl2012,AdessoRagyLee2014}.
Consequently, the presence of negative eigenvalues, and hence the
CP-divisibility classification, is invariant, although their numerical
magnitudes are generally frame dependent.\\

CP-divisibility characterizes whether the reduced evolution between
any two times admits a CPTP intermediate map. Its violation establishes
non-Markovianity in the divisibility sense, but does not by itself imply
operational information backflow, which requires an independent revival
of an information measure, such as the distinguishability of suitably
chosen reduced states
\cite{BreuerLainePiilo2009,RivasHuelgaPlenio2014,BreuerEtAl2016}.
Moreover, CP-divisibility constrains the reduced two-time maps but does
not exclude memory stored in higher-order temporal correlations; a
CP-divisible evolution may therefore still exhibit genuine multitime
non-Markovianity \cite{MilzKimPollockModi2019}.

\subsection{BLP: recovery of preparation distinguishability}
A complementary operational criterion is provided by the
Breuer--Laine--Piilo (BLP) approach, which characterizes information
backflow through the distinguishability of two system preparations
evolving under the same reduced dynamical map
\cite{BreuerLainePiilo2009,LainePiiloBreuer2010,BreuerEtAl2016}.
For two initial curvature states
$\hat\rho_{\mathcal S,k}^{(i)}(\eta_0)$, $i=1,2$, their evolved trace
distance at time $\eta$ is 
\begin{eqnarray}
D_k(\eta)
&\equiv&
\frac12
\left\|
\hat\rho_{\mathcal S,k}^{(1)}(\eta)
-
\hat\rho_{\mathcal S,k}^{(2)}(\eta)
\right\|_1 
\label{eq:trace_distance}
\end{eqnarray}
where
$\|\hat A\|\equiv
{\rm Tr}\sqrt{\hat A^\dagger\hat A}$
denotes the trace norm. 
When the two initial curvature preparations are assumed to occur with equal prior probabilities, $p_1=p_2=1/2$, the trace distance determines the optimal single-shot discrimination probability,
$P_{\rm succ}(\eta)=[1+D_k(\eta)]/2$
\cite{Helstrom1969}.
Since the trace distance is contractive under any CPTP map,
$D[\Phi(\rho_1),\Phi(\rho_2)]\leqslant D(\rho_1,\rho_2)$, CP-divisible dynamics requires $D_k(\eta)$ to be nonincreasing for every pair of initial states. Hence, the existence
of a pair for which
$\dd D_k/\dd\eta \geqslant 0$ provides an operational witness of information
backflow and, consequently, of CP-indivisibility
\cite{BreuerLainePiilo2009,RivasHuelgaPlenio2014}.
The corresponding BLP measure is
\begin{eqnarray}
\mathcal N_{{\rm BLP},k}
&=&
\sup_{\rho_{1,2}(\eta_0)}
\int_{D_k'\geqslant 0}
\dd\eta\,D_k'(\eta),
\qquad
D_k'
\equiv
\frac{\dd D_k}{\dd\eta}.
\label{eq:blp_measure}
\end{eqnarray}
The analysis below evaluates selected pairs of preparations and
therefore provides BLP witnesses rather than the global optimization
in Eq.~(\ref{eq:blp_measure}). Conversely, CP-indivisibility need not
produce a trace-distance revival for a given pair, and the RHP and BLP
criteria are therefore not equivalent in general
\cite{RivasHuelgaPlenio2014,BreuerEtAl2016}.\\

For the Gaussian dynamics considered here, we choose two families of
curvature probes oriented along the $q$ and $p$ quadratures. Within
each family, the two preparations share the same initial covariance
$\boldsymbol{\sigma}_{\mathcal S,k}(\eta_0)$ and have opposite first moments,
\begin{eqnarray}
\mathbf d_{\mathcal S,k}^{(1)}(\eta_0)
&=&
-\mathbf d_{\mathcal S,k}^{(2)}(\eta_0)
=
\boldsymbol{\delta}_{j},
\qquad
\boldsymbol{\delta}_{j}
=
\frac{\mathbf e_j}{
\sqrt{
\mathbf e_j^T
\boldsymbol{\sigma}_{\mathcal S,k}^{-1}(\eta_0) 
\mathbf e_j
}},
\qquad
j\in\{q,p\},
\label{eq:gaussian_pair}
\end{eqnarray}
where $\mathbf e_q=(1,0)^T$ and $\mathbf e_p=(0,1)^T$ 
define the
$q$- and $p$-oriented 
directions in curvature 
phase space,  respectively.  
The normalization
$\boldsymbol{\delta}_{j}^{T}
\boldsymbol{\sigma}_{\mathcal S,k}^{-1}(\eta_0)
\boldsymbol{\delta}_{j}=1$
for each $j\in{q,p}$ fixes the same initial covariance-normalized
displacement for the two probe orientations, placing them on an equal
footing even when the initial covariance is anisotropic 
owing to unequal variances of
phase-space fluctuations
along different directions. 
Because the two preparations evolve under the same Gaussian channel,
their covariance matrices remain identical and hence have the same
purity. Their first-moment separation evolves as
$
\Delta\mathbf d_{\mathcal S,k}(\eta)
 \equiv 
\mathbf d_{\mathcal S,k}^{(1)}(\eta)
-
\mathbf d_{\mathcal S,k}^{(2)}(\eta)
=
2X_k(\eta,\eta_0)\boldsymbol{\delta}_{j}$,
where the common channel displacement cancels in the difference.
For two Gaussian states 
$\hat\rho_{\mathcal S,k}^{(1,2)}$, with identical covariance matrix
$\boldsymbol{\sigma}_{\mathcal S,k}$, the squared Uhlmann fidelity can be written directly in terms of their first-moment
separation and common covariance matrix
\cite{BanchiBraunsteinPirandola2015} as 
\begin{eqnarray}
{\rm Fid}_k^{(\bm{\delta}_j)} (\eta)
&\equiv&
\left[
{\rm Tr}
\sqrt{
\sqrt{\hat\rho_{\mathcal S,k}^{(1)}}
\hat\rho_{\mathcal S,k}^{(2)}
\sqrt{\hat\rho_{\mathcal S,k}^{(1)}}
}
\right]^2\nonumber\\
&=&  
\exp\!\left[
-\frac14
\Delta\mathbf d_{\mathcal S,k}^{T}
\boldsymbol{\sigma}_{\mathcal S,k}^{-1}
\Delta\mathbf d_{\mathcal S,k}
\right] 
=
\exp\!\left[
-\bm{\delta}_j^T X_k^T(\eta,\eta_0) 
\boldsymbol{\sigma}_{\mathcal S,k}^{-1}(\eta)
X_k(\eta,\eta_0) \bm{\delta}_j 
\right].
\label{eq:fid}
\end{eqnarray}
At $\eta=\eta_0$, 
 both probe orientations $j\in \{q,p\}$ have the same
initial fidelity 
${\rm Fid}_k^{(\bm{\delta}_j)} (\eta_0)=e^{-1}$
owing to the normalization 
$\boldsymbol{\delta}_j^{T}
\boldsymbol{\sigma}_{\mathcal S,k}^{-1}(\eta_0)
\boldsymbol{\delta}_j=1$.
For pure initial probe states, the squared Uhlmann fidelity and trace
distance satisfy
$D_k(\eta_0)=\sqrt{1-{\rm Fid}_k(\eta_0)}$
\cite{WeedbrookEtAl2012,BanchiBraunsteinPirandola2015}, yielding $
D_k(\eta_0) = \sqrt{1-e^{-1}}
\simeq 0.795$ 
for both the $q$- and $p$-oriented pairs.\\

Although a curvature-mode Gaussian state is fully specified by its first
moments and $2\times2$ covariance matrix, its density operator acts on
the infinite-dimensional bosonic Fock space. Consequently, no general
simple closed-form expression for the trace distance between two mixed
Gaussian states is available directly in terms of their moments
\cite{WeedbrookEtAl2012,MartinezCifuentesQuesada2026}. We therefore
reconstruct each evolved Gaussian state in the Fock basis from its first
moments and covariance matrix
\cite{DodonovMankoManko1994,QuesadaEtAl2019}. Truncating the basis at
occupation $n_{\rm cut}$ yields finite-dimensional approximations
$\rho_{1,2}^{(n_{\rm cut})}$, from which
\begin{eqnarray}
D_k
&=&
\frac12
{\rm Tr}
\left|
\hat\rho_1^{(n_{\rm cut})}
-
\hat\rho_2^{(n_{\rm cut})}
\right|
=
\frac12
\sum_{\ell}
|\lambda_\ell|, 
\label{eq:fock_trace_distance}
\end{eqnarray}
where $\lambda_\ell$ are the eigenvalues of
$\hat\rho_1^{(n_{\rm cut})}-\hat\rho_2^{(n_{\rm cut})}$.
Numerical convergence is verified by increasing $n_{\rm cut}$ and
monitoring both $D_k$ and the discarded occupation weight,
$1-{\rm Tr}\hat\rho_i^{(n_{\rm cut})}$. 
 This construction evaluates the trace distance directly, avoiding its replacement by Gaussian fidelity, which constitutes a distinct distinguishability measure \cite{BanchiBraunsteinPirandola2015}. The displaced probe pairs are used only as controlled diagnostics of the reduced channel and are not identified with physical primordial-vacuum preparations.

\section{Quantum Fisher information and metrological recovery}
\label{sec:qfi}

\subsection{Parameter encoding and Gaussian QFI} 
To probe the recovery of metrological sensitivity initially encoded in
the curvature mode, we introduce a parameter $\vartheta$ exclusively
through the initial curvature preparation, while keeping the initial
entropic state and the subsequent reduced dynamics independent of
$\vartheta$. Thus,
$\partial_\vartheta\Phi_k(\eta,\eta_0) = 0$,
or equivalently
$\partial_\vartheta X_k
=\partial_\vartheta Y_k
=\partial_\vartheta c_k=0$.
The quantum Fisher information (QFI) associated with estimating
$\vartheta$ from the evolved reduced curvature state
$\hat\rho_{\mathcal S,k}(\eta;\vartheta)$ is
\begin{eqnarray}
F_{Q,k}(\eta;\vartheta)
&\equiv&
{\rm Tr}\!\left[
\hat\rho_{\mathcal S,k}(\eta;\vartheta)
\hat L_{\vartheta,k}^{\,2}(\eta;\vartheta)
\right],
\label{eq:curvature_QFI_definition}
\end{eqnarray}
where the symmetric logarithmic derivative
$\hat L_{\vartheta,k}$ is defined by 
$2\partial_\vartheta\hat\rho_{\mathcal S,k}
=
\{
\hat\rho_{\mathcal S,k},
\hat L_{\vartheta,k} \}$.
For $n$ independently prepared copies, any locally unbiased estimator
$\widetilde{\vartheta}$ obeys the quantum Cram\'er--Rao bound
${\rm Var}(\widetilde{\vartheta})\geqslant[nF_{Q,k}]^{-1}$, so that
$F_{Q,k}$ quantifies the optimal local sensitivity to the encoded
parameter
\cite{Helstrom1967,BraunsteinCaves1994,Paris2009}. 
For a general single-mode Gaussian state, the QFI is determined by its
first moments $\mathbf d_{\mathcal S,k}$, covariance matrix
$\boldsymbol{\sigma}_{\mathcal S,k}$, and purity
$\mu_{\mathcal S,k}$ as
\cite{PinelEtAl2013,Safranek2019}
\begin{eqnarray}
F_{Q,k}
&=&
\left(
\partial_\vartheta\mathbf d_{\mathcal S,k}
\right)^T
\boldsymbol{\sigma}_{\mathcal S,k}^{-1}
\left(
\partial_\vartheta\mathbf d_{\mathcal S,k}
\right) 
+
\frac{
{\rm Tr}\!\left[
\left(
\boldsymbol{\sigma}_{\mathcal S,k}^{-1}
\partial_\vartheta\boldsymbol{\sigma}_{\mathcal S,k}
\right)^2
\right]
}{
2\left(1+\mu_{\mathcal S,k}^{\,2}\right)
}
+
\frac{
2\left(
\partial_\vartheta\mu_{\mathcal S,k}
\right)^2
}{
1-\mu_{\mathcal S,k}^{\,4}
}.
\label{eq:gaussian_qfi}
\end{eqnarray}
 For pure Gaussian states, $\mu_{\mathcal S,k}=1$, the formally singular purity term in Eq.~(\ref{eq:gaussian_qfi}) is evaluated through the corresponding pure-state limit \cite{Safranek2019}.
Since the reduced channel is $\vartheta$ independent,
Eqs.~(\ref{eq:reduced_first_moments}) and
(\ref{eq:reduced_covariance_channel}) give
\begin{eqnarray}
\partial_\vartheta\mathbf d_{\mathcal S,k}(\eta)
&=&
X_k(\eta,\eta_0)
\partial_\vartheta\mathbf d_{\mathcal S,k}(\eta_0),
\nonumber\\
\partial_\vartheta\boldsymbol{\sigma}_{\mathcal S,k}(\eta)
&=&
X_k(\eta,\eta_0)
\left[
\partial_\vartheta
\boldsymbol{\sigma}_{\mathcal S,k}(\eta_0)
\right]
X_k^T(\eta,\eta_0).
\end{eqnarray}
We consider a displacement-only encoding of the initial curvature
state,
\begin{eqnarray}
\mathbf d_{\mathcal S,k}(\eta_0;\vartheta)
&=&
\mathbf d_{\mathcal S,k}(\eta_0;0)
+
\vartheta\,\mathbf u_k,
\qquad
\partial_\vartheta
\boldsymbol{\sigma}_{\mathcal S,k}(\eta_0)
=
0,
\label{eq:displacement_encoding}
\end{eqnarray}
where $\mathbf u_k$ specifies the encoding direction in curvature phase
space. The covariance matrix and purity are then independent of
$\vartheta$, and the QFI reduces to
\begin{eqnarray}
F_{Q,k}^{(\mathbf u)}(\eta)
&=&
\mathbf u_k^T
X_k^T(\eta,\eta_0)
\boldsymbol{\sigma}_{\mathcal S,k}^{-1}(\eta)
X_k(\eta,\eta_0)
\mathbf u_k .
\label{eq:displacement_qfi}
\end{eqnarray}
For direct comparison with the BLP probes, we choose
$\mathbf u_k=\boldsymbol{\delta}_q$ or
$\boldsymbol{\delta}_p$ from Eq.~(\ref{eq:gaussian_pair}).
Their normalization gives
$F_{Q,k}^{(\mathbf u)}(\eta_0)=1$ for either orientation, so that subsequent
variations directly quantify the loss and recovery of the initially
encoded sensitivity. The parameter $\vartheta$ is therefore an
auxiliary probe parameter describing an initial curvature displacement.\\

During inflation, the curvature fluctuations generally develop unequal variances along the $q$ and $p$ directions, rendering the covariance anisotropic. Consequently, the QFI associated with displacement encodings along the two quadratures may evolve differently. This effect should be distinguished from unitary squeezing: the QFI is invariant under any $\vartheta$-independent unitary transformation, so parameter-independent squeezing of an isolated curvature mode cannot by itself modify the QFI \cite{BraunsteinCaves1994,Paris2009}. In the present reduced description, temporal variations of $F_{Q,k}$ therefore arise from the nonunitary part of the curvature dynamics induced by coupling to the entropic sector.

\subsection{Data processing and comparison with trace distance}

For any $\vartheta$-independent CPTP map $\Lambda$, the QFI obeys the
data-processing inequality
\cite{BraunsteinCaves1994,PetzSudar1996,Paris2009}
\begin{eqnarray}
F_Q\!\left(
\Phi[\hat\rho_\vartheta]
\right)
&\leqslant&
F_Q(\hat\rho_\vartheta),
\qquad
\partial_\vartheta\Phi=0 .
\label{eq:qfi_data_processing}
\end{eqnarray}
Consequently, for the fixed-initial-time channel considered here, $F_{Q,k}(\eta)
 \leqslant 
F_{Q,k}(\eta_0)$,
irrespective of whether the reduced evolution is Markovian. This bound
does not preclude a later recovery of QFI following an earlier loss; it
only prevents the recovered sensitivity from exceeding its initial
value under the same CPTP map.
If the dynamics is CP divisible, every intermediate map is CPTP and
parameter independent, implying the stronger monotonicity condition
$F_{Q,k}(\eta_2) \leqslant F_{Q,k}(\eta_1)$ for
$\eta_2\geqslant \eta_1$. 
A positive QFI flow,
\begin{eqnarray}
\mathcal J_{Q,k}(\eta)
&\equiv&
\frac{\partial F_{Q,k}}{\partial\eta}
\geqslant 0 ,
\label{eq:qfi_flow}
\end{eqnarray}
therefore witnesses CP-indivisibility wherever the derivative exists
\cite{LuWangSun2010,AbiusoEtAl2023}. The converse need not hold for a
fixed input state or parameter encoding, so the QFI flow provides an
operational witness rather than a complete characterization of
CP-indivisibility.\\

For the choice
$\mathbf u_k=\boldsymbol{\delta}_q$ or
$\boldsymbol{\delta}_p$, introduced above to compare the QFI with the
equal-covariance, oppositely displaced BLP probes of
Eq.~(\ref{eq:gaussian_pair}), the squared Uhlmann fidelity in
Eq.~(\ref{eq:fid}) can be related directly to the displacement QFI.
Using Eq.~(\ref{eq:displacement_qfi}), one finds
\begin{eqnarray}
{\rm Fid}_k^{(\boldsymbol{\delta}_j)}(\eta)
&=&
\exp\!\left[
-
F_{Q,k}^{(\boldsymbol{\delta}_j)}(\eta)
\right],
\qquad
j\in\{q,p\}.
\label{eq:fid1}
\end{eqnarray}
Thus, for these displacement probes, the QFI directly determines the
fidelity between the two evolved preparations. 
Once the reduced curvature probes become mixed during the evolution,
the trace distance is no longer uniquely determined by the fidelity.
For the squared Uhlmann fidelity adopted here, the two quantities obey
the Fuchs--van de Graaf inequalities
\cite{FuchsVanDeGraaf1999},
\begin{eqnarray}
1-\sqrt{{\rm Fid}_k^{(\boldsymbol{\delta}_j)}(\eta)}
&\leqslant&
D_k(\eta)
\leqslant
\sqrt{1-{\rm Fid}_k^{(\boldsymbol{\delta}_j)}(\eta)} .
\label{eq:fuchs_van_de_graaf}
\end{eqnarray}
Thus, once the reduced probes become mixed, the QFI constrains their
distinguishability but no longer uniquely determines the trace distance.
For pure states, the upper bound is saturated,
yielding
$D_k=\sqrt{1-{\rm Fid}_k^{(\boldsymbol{\delta}_j)}}$, whereas this
relation does not hold in general for mixed states
\cite{FuchsVanDeGraaf1999,WeedbrookEtAl2012},
for which  the
trace distance additionally 
depends on the common Gaussian covariance matrix.
 Mixedness of Gaussian curvature states is 
characterized by the symplectic
eigenvalue of the covariance matrix,
$\nu_{\mathcal S,k} = \sqrt{
\det\boldsymbol{\sigma}_{\mathcal S,k}
}$, which measures the phase-space area 
occupied by the Gaussian state and
therefore its departure from minimum uncertainty. In the covariance
convention adopted here, the uncertainty relation implies
$\nu_{\mathcal S,k}\geqslant 1/2$, with
$\nu_{\mathcal S,k}=1/2$ for a pure Gaussian state and
$\nu_{\mathcal S,k}\geqslant 1/2$ for a mixed one
\cite{WeedbrookEtAl2012,AdessoRagyLee2014}. \\

Consequently, a revival of the trace distance for the selected BLP pair
need not coincide with a positive QFI flow, although either provides a
witness of CP-indivisibility under the parameter-independent channel
assumption. The two quantities therefore probe complementary aspects
of information recovery: the trace distance tracks recovery of state
distinguishability, whereas the QFI tracks recovery of sensitivity to
the encoded parameter. By contrast, an increase in the reduced purity
$\mu_{\mathcal S,k}$ indicates reduced mixedness and is not tied to a
specific preparation pair or parameter encoding.

\section{Numerical methodology}
\label{sec:numerical_methodology}

We now evaluate the reduced curvature dynamics numerically using a
prescribed inflation-inspired background. The purpose is to compare,
within the same quadratic curvature--entropic evolution, reduced-state
recoherence, CP-indivisibility, recovery of preparation
distinguishability, and recovery of displacement-encoded QFI. The
background quantities are specified phenomenologically rather than
obtained from a particular two-field potential, so that the analysis
isolates the quantum-information dynamics without implying
model-specific predictions for cosmological observables.

\subsection{Background and interaction profile}  
\label{subsec:nu1}
We use the number of e-folds measured relative to the epoch of maximum
turning rate of the homogeneous background trajectory in scalar-field
space,
\begin{eqnarray}
N
&\equiv&
\ln\!\left(\frac{a}{a_t}\right),
\qquad
a_t\equiv a(N_t),
\end{eqnarray}
as the numerical time variable. Choosing $N_t=0$ and normalizing
$a_t=1$, a constant slow-roll parameter
$\epsilon\equiv-\dot H/H^2$ gives
\begin{eqnarray}
a(N)
&=&
e^N,
\qquad
H(N)
=
H_t e^{-\epsilon N},
\label{eq:numerical_background}
\end{eqnarray}
where $H_t\equiv H(N_t)=H(0)$. For canonical scalar fields,
$z=a\sqrt{2\epsilon}\,M_{\rm Pl}$, so that for constant $\epsilon$
\begin{eqnarray}
\frac{z'}{z}
&=&
aH,
\qquad
\frac{z''}{z}
=
\frac{a''}{a}
=
(2-\epsilon)(aH)^2 .
\label{eq:numerical_background_derivatives}
\end{eqnarray}
The entropic mass parameter $M_{\mathcal F}$ defined in
Eq.~(\ref{eq:entropic_mass_coupling}) is specified through a constant
ratio $M_{\mathcal F}/H$. The dimensionless turn rate
$\alpha=\dot\theta/H$, introduced in
Eq.~(\ref{eq:lambda_definition}), is prescribed by
\begin{eqnarray}
\alpha(N)
&=&
\alpha_0
\exp\!\left(
-\frac{N^2}{2\sigma_N^2}
\right),
\qquad
\lambda(N)
=
2aH\,\alpha(N),
\label{eq:numerical_turn_profile}
\end{eqnarray}
where $\alpha_0$ is the maximum turn rate and $\sigma_N$ sets its width
in e-folds. 
The Gaussian profile describes a localized turn of the homogeneous background trajectory in scalar-field space, for which the turning rate is significant only over a finite interval in e-fold time. This provides a controlled curvature--entropic interaction window and allows modes encountering the interaction at different stages 
relative to Hubble crossing to be compared 
within the same background \cite{ColasGrainVennin2023,KranasGrainVennin2025}.  
We characterize the Fourier scale relative to the Hubble
scale at the peak of the turn by the dimensionless parameter
\begin{eqnarray}
\kappa_t
&\equiv&
\left.
\frac{k}{aH}
\right|_{N=0}
=
\frac{k}{H_t}.
\label{eq:kappa_turn}
\end{eqnarray}
Using Eq.~(\ref{eq:numerical_background}), the instantaneous
ratio of the physical wave number to the Hubble scale is
given by
\begin{eqnarray}
\frac{k}{aH}
&=&
\kappa_t e^{-(1-\epsilon)N}.
\label{eq:kappa_evolution}
\end{eqnarray}
While $k/(aH)$ characterizes the instantaneous Hubble-scale
regime of a Fourier mode, $\kappa_t$ specifies its regime
at the peak of the turn.
Accordingly, $\kappa_t\ll1$, $\kappa_t\simeq1$, and
$\kappa_t\gg1$ correspond, respectively, to modes that
are super-Hubble, near Hubble crossing, and sub-Hubble
at this epoch.
Denoting the epoch of Hubble crossing by $N_k$, defined
through $k=a(N_k)H(N_k)$, one obtains
\begin{eqnarray}
N_k
&=&
\frac{\ln\kappa_t}{1-\epsilon}.
\label{eq:hubble_crossing_epoch}
\end{eqnarray}
Thus, $\kappa_t$ also determines the temporal separation
between Hubble crossing and the peak of the turn, allowing
the reduced curvature dynamics to be investigated for
modes encountering the curvature--entropic interaction
at different stages of their cosmological evolution
within the same inflationary background.
All
dimensionful quantities are expressed in units of $H_t$, with $H_t=1$. For the numerical analysis, the dynamics is specified by the slow-roll
parameter $\epsilon$, the entropic mass ratio $M_{\mathcal F}/H$, the
maximum turn rate $\alpha_0$, the turn width $\sigma_N$, and the Fourier-scale
parameter $\kappa_t$. 
To define the reference evolutions used in the numerical analysis, we fix
the background and interaction parameters to
\begin{eqnarray}
\epsilon = 0.01,
\qquad
\frac{M_{\mathcal F}}{H}=2,
\qquad
\alpha_0=1.0,
\qquad
\sigma_N = 1,
\label{eq:benchmark_parameters}
\end{eqnarray}
and evolve the system over $N\in[N_i=-8,N_f=3]$. For the
Fourier-scale comparison, we consider
\begin{eqnarray}
\kappa_t \in \{0.1,\,1,\,1.0\},
\end{eqnarray}
while keeping the remaining parameters fixed at the reference values
given in Eq.~(\ref{eq:benchmark_parameters}). In the parameter scans and
other scale-dependent studies, the parameter(s) of interest are varied
over the ranges specified for the corresponding figures, with all
remaining quantities held fixed at their reference values unless stated
otherwise.

\subsection{Initial state and symplectic evolution}
\label{subsec:nu2}
At $N=N_i$, the curvature and entropic sectors are initialized in a
factorized, centered, pure Gaussian state approaching the
Bunch--Davies vacuum \cite{BunchDavies1978}, with the initial time
chosen sufficiently before the localized turn such that
$\alpha(N_i)\simeq0$ and hence the curvature--entropic coupling is
negligible.  
For numerical conditioning, we perform a constant symplectic rescaling
of the joint curvature--entropic phase-space variables,
\begin{eqnarray}
\widetilde{\mathbf R}_k
&=&
S_k\mathbf R_k,
\qquad
S_k
=
{\rm diag}
\left(
\sqrt{k},
\frac{1}{\sqrt{k}},
\sqrt{k},
\frac{1}{\sqrt{k}}
\right),
\qquad
S_k\Omega S_k^T=\Omega ,
\label{eq:numerical_symplectic_scaling}
\end{eqnarray}
under which the initial covariance matrix $\boldsymbol{\sigma}_k(N_i)$
is transformed to 
$\widetilde{\boldsymbol{\sigma}}_k(N_i) =
\frac12\mathbb I_4$.
This rescaling changes only the phase-space representation and leaves
the relevant symplectic invariants and the CP-divisibility
classification unchanged.
Using $\dd/\dd N=(aH)^{-1}\dd/\dd\eta$, the rescaled propagator
$\widetilde M_k=S_kM_kS_k^{-1}$ obeys
\begin{eqnarray}
\frac{\dd \widetilde M_k}{\dd N}
&=&
\widetilde{\mathcal A}_k(N)\widetilde M_k,
\qquad
\widetilde{\mathcal A}_k(N)
=
S_k\frac{\Omega G_k(N)}{aH}S_k^{-1},
\qquad
\widetilde M_k(N_i)=\mathbb I_4 ,
\label{eq:numerical_propagator}
\end{eqnarray}
where $G_k$ is given in Eq.~(\ref{eq:G_matrix}).  
The evolution is integrated using the adaptive eighth-order
Dormand--Prince DOP853 scheme
\cite{PrinceDormand1981,HairerNorsettWanner1993}.  
 The
joint covariance matrix and the reduced-channel matrices $X_k$ and
$Y_k$ are then obtained from
Eqs.~(\ref{eq:exact_gaussian_evolution}) and
(\ref{eq:reduced_covariance_channel}). Since the initial entropic state
is centered, $\mathbf d_{\mathcal E,k}(N_i)=0$, the channel displacement
vanishes, $c_k=0$. 
Numerical accuracy is monitored through the symplecticity error
\begin{eqnarray}
\epsilon_{\rm symp}(N)
&\equiv&
\left\|
\widetilde M_k\Omega\widetilde M_k^T-\Omega
\right\|_F ,
\label{eq:symplectic_error}
\end{eqnarray}
where $\|A\|_F\equiv\sqrt{{\rm Tr}(A^\dagger A)}$ denotes the
Frobenius norm. Additional consistency checks include conservation of
the joint purity, the Robertson--Schr\"odinger physicality conditions
for the joint and reduced covariance matrices, and the
Gaussian-channel complete-positivity condition in
Eq.~(\ref{eq:gaussian_channel_CP}).
 For the benchmark evolution, the adaptive solver is run with relative and absolute error tolerances of $10^{-12}$ and $10^{-14}$, respectively, controlling the local truncation error relative to the solution magnitude and in absolute terms, with a maximum step size $\Delta N=0.02$.

\subsection{Numerical diagnostics}
\label{subsec:nu3}
The reduced curvature purity is evaluated from
Eq.~(\ref{eq:gaussian_purity}). For the initially pure reduced state, we
quantify the fraction of purity recovered after its minimum by
\begin{eqnarray}
\mathcal R_\mu
&\equiv&
\frac{
\mu_{\mathcal S,k}(N_f)-\mu_{\mathcal S,k}^{\rm min}
}{
1-\mu_{\mathcal S,k}^{\rm min}
},
\qquad
\mu_{\mathcal S,k}^{\rm min}
\equiv
\min_{N\in[N_i,N_f]}
\mu_{\mathcal S,k}(N),
\label{eq:recoherence_fraction}
\end{eqnarray}
such that $\mathcal R_\mu=0$ and $1$ correspond, respectively, to no
recovery and complete recovery of the lost purity.\\

CP-divisibility is tested using the finite-interval Gaussian CP matrix
$C_{21}$ of Eq.~(\ref{eq:intermediate_cp}) between successive numerical
times. CP-indivisibility is identified when the smallest eigenvalue of
the finite-interval CP matrix satisfies
$\lambda_{\min}(C_{21}) \leqslant -10^{-8}$, where the threshold is introduced to
exclude numerical roundoff. The local matrix $C_{\rm rate}$ in
Eq.~(\ref{eq:cp_rate}) is evaluated independently as a consistency
check; since the magnitudes of its eigenvalues depend on the
phase-space frame, only their signs are used in the divisibility
classification. 
For visualization, we introduce the binary RHP indicator
\begin{eqnarray}
\mathcal I_{\rm RHP}(N)
=
\begin{cases}
1, & \lambda_{\min}(C_{21}) \leqslant -10^{-8},\\
0, & \lambda_{\min}(C_{21})\geqslant -10^{-8},
\end{cases}
\label{eq:rhp_binary_indicator}
\end{eqnarray}
so that $\mathcal I_{\rm RHP}=1$ identifies a CP-indivisible
intermediate map, whereas $\mathcal I_{\rm RHP}=0$ denotes no detected
violation of CP-divisibility at the adopted numerical threshold. This
binary indicator is used in the numerical plots in
Sec.\ \ref{sec:numerical_results}.\\

For the BLP diagnostic, we propagate the equal-covariance,
oppositely displaced $q$- and $p$-oriented probe pairs defined in
Eq.~(\ref{eq:gaussian_pair}).
At each time, the common covariance matrix of the two probes is brought
to Williamson normal form by a common Gaussian unitary
\cite{WeedbrookEtAl2012,AdessoRagyLee2014}. By unitary invariance of
the trace distance, the calculation is then reduced to two oppositely
displaced Gaussian states with covariance
$\nu_{\mathcal S,k}\mathbb I_2$, where
$\nu_{\mathcal S,k}=\sqrt{\det\boldsymbol{\sigma}_{\mathcal S,k}}$;
equivalently, this covariance corresponds to a thermal state with mean
occupation $\bar n_k=\nu_{\mathcal S,k}-1/2$. 
  The corresponding density operators
are reconstructed in a truncated Fock basis
\cite{DodonovMankoManko1994,QuesadaEtAl2019}, and the trace distance is
evaluated directly from Eq.~(\ref{eq:trace_distance}).  
The Fock cutoff $n_{\rm cut}$ introduced in
Eq.~(\ref{eq:fock_trace_distance}) is increased until the discarded
occupation weight is below $10^{-10}$ and the resulting change in
$D_k$ is below $10^{-8}$.\\

For the same probe directions, the displacement QFI is evaluated
directly from Eq.~(\ref{eq:displacement_qfi}), without Fock-space
reconstruction.  As independent consistency checks, we verify
Eq.~(\ref{eq:fid1}) relating the displacement QFI to the squared
Uhlmann fidelity and the Fuchs--van de Graaf bounds in
Eq.~(\ref{eq:fuchs_van_de_graaf}). 
To quantify the accumulated revival of distinguishability and QFI for
the selected probes, we define
\begin{eqnarray}
\mathcal B_j
&\equiv&
\sum_n
\left[
D_j(N_{n+1})-D_j(N_n)
\right]_+,
\qquad
\mathcal Q_j
\equiv
\sum_n
\left[
F_{Q,k}^{(\boldsymbol{\delta}_j)}(N_{n+1})
-
F_{Q,k}^{(\boldsymbol{\delta}_j)}(N_n)
\right]_+ ,
\label{eq:numerical_positive_increments}
\end{eqnarray}
where $j\in\{q,p\}$, $[x]_+\equiv\max(x,0)$, and $D_j$ denotes the
trace distance for the $j$-oriented probe pair. These quantities
accumulate all positive-flow intervals and are therefore sensitive to
repeated revivals during the evolution. In particular, $\mathcal B_j$
is a selected-probe quantity and should not be identified with the
globally optimized BLP measure in Eq.~(\ref{eq:blp_measure}). 
To complement the accumulated revival measures, the fractional recovery from the maximum suppression is defined as 
\begin{eqnarray}
\mathcal R_{D,j}
&\equiv&
\frac{
D_j(N_f)-D_j^{\rm min}
}{
D_j(N_i)-D_j^{\rm min}
},
\qquad
D_j^{\rm min}
\equiv
\min_{N\in[N_i,N_f]}D_j(N),
\nonumber\\
\mathcal R_{Q,j}
&\equiv&
\frac{
F_{Q,k}^{(\boldsymbol{\delta}_j)}(N_f)
-
F_{Q,k}^{(\boldsymbol{\delta}_j),{\rm min}}
}{
F_{Q,k}^{(\boldsymbol{\delta}_j)}(N_i)
-
F_{Q,k}^{(\boldsymbol{\delta}_j),{\rm min}}
},
\qquad
F_{Q,k}^{(\boldsymbol{\delta}_j),{\rm min}}
\equiv
\min_{N\in[N_i,N_f]}
F_{Q,k}^{(\boldsymbol{\delta}_j)}(N).
\label{eq:operational_recovery_fraction}
\end{eqnarray}
The two sets of measures provide complementary characterizations of
information recovery. While $\mathcal B_j$ and $\mathcal Q_j$ quantify
the accumulated positive variations of distinguishability and
displacement QFI throughout the evolution, $\mathcal R_{D,j}$ and
$\mathcal R_{Q,j}$ measure the fractions of their respective maximum
losses recovered by the final time. 
These recovery fractions are
well defined when the corresponding suppression relative to the initial
value is nonzero; when this suppression becomes sufficiently small,
their numerical evaluation becomes increasingly sensitive.

\section{Numerical results and discussion}
\label{sec:numerical_results}  
\subsection{Fourier-scale dependence of the reduced information dynamics}

 \begin{figure}[!h]
\centering
\includegraphics[width=0.98\textwidth]{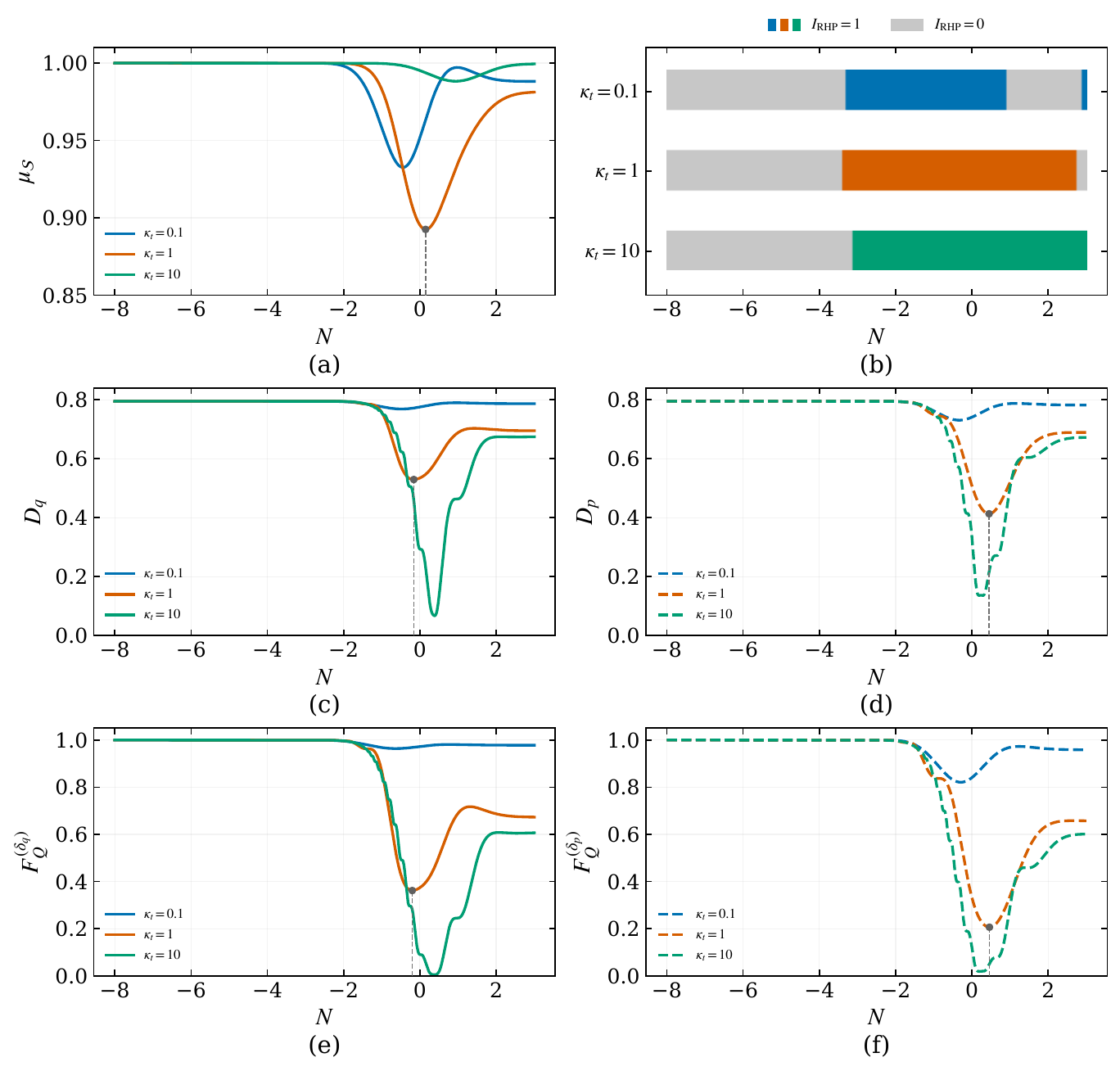}
\caption{Scale-resolved dynamics of the reduced curvature mode for
$\epsilon=0.01$, $M_{\mathcal F}/H=2$, $\alpha_0=1$, and $\sigma_N=1$.
At the peak of the turn ($N=0$), the modes with $\kappa_t=0.1$, $1$, and
$10$ are super-Hubble, at Hubble crossing, and sub-Hubble, respectively.
Panel (a) shows the reduced curvature purity, and panel (b) shows the
finite-interval RHP CP-indivisibility indicator. Panels (c) and (d) show
the trace distances for the $q$- and $p$-oriented probe pairs, respectively,
and panels (e) and (f) show the corresponding displacement QFI.
Blue, orange, and green denote $\kappa_t=0.1$, $1$, and $10$, respectively.
In panels (c)--(f), solid curves denote $q$ displacements and dashed
curves denote $p$ displacements. The full-height dotted line marks $N=0$
in every panel. In panels (a) and (c)--(f), grey dots mark the sampled
minima of the $\kappa_t=1$ curves over $-8\leqslant N\leqslant3$, and short
vertical dashed lines mark the e-fold at which each minimum occurs.}

\label{fig:three_kappa_refined}
\end{figure}

Figure~\ref{fig:three_kappa_refined} shows the evolution of four information-theoretic diagnostics as functions of the e-folding parameter $N$: the reduced curvature purity $\mu_{\mathcal S,k}$, the finite-interval RHP CP-indivisibility indicator ($\mathcal I_{\rm RHP}$  in Eq.\ \eqref{eq:rhp_binary_indicator}), the trace distances $D_q$ and $D_p$ for the selected $q$- and $p$-oriented probe pairs, and the corresponding displacement QFI, $F_{Q,k}^{(\delta_q)}$ and $F_{Q,k}^{(\delta_p)}$. The evolution is shown for the representative values $\kappa_t=0.1,\,1,$ and $10$, corresponding, respectively, to modes that are super-Hubble, near Hubble crossing, and sub-Hubble at the peak of the turn, with all remaining parameters fixed at the reference values given in Eq.~(\ref{eq:benchmark_parameters}).  
For a quantitative comparison, Table~\ref{tab:kappa_scale_minima}
lists the minima of the state- and probe-dependent diagnostics
and the corresponding e-folds at which they occur.
Table~\ref{tab:kappa_scale_recovery} summarizes the associated
recovery and revival measures, including the purity-recovery
fraction $\mathcal R_\mu$, the normalized trace-distance and
QFI recovery fractions $\mathcal R_{D,j}$ and $\mathcal R_{Q,j}$
defined in Eq.~(\ref{eq:operational_recovery_fraction}),
and the accumulated positive increments $\mathcal B_j$
and $\mathcal Q_j$ defined in
Eq.~(\ref{eq:numerical_positive_increments}).
This comparison elucidates the Fourier-scale dependence of
the distinct information-theoretic diagnostics characterizing
the reduced curvature dynamics.\\

\begin{table*}[t]
\centering
\renewcommand{\arraystretch}{1.25}
\begin{tabular}{c|c|cc|cc} 
\hline\hline
$\kappa_t$
&
$\mu_{\mathcal S,k}^{\rm min}\;(N_{\mu}^{\rm min})$
&
$D_q^{\rm min}\;(N_{D_q}^{\rm min})$
&
$D_p^{\rm min}\;(N_{D_p}^{\rm min})$
&
$F_{Q,k}^{(\delta_q),{\rm min}}\;(N_{F_q}^{\rm min})$
&
$F_{Q,k}^{(\delta_p),{\rm min}}\;(N_{F_p}^{\rm min})$
\\
\hline
$0.1$
&
$0.933\;(-0.44)$
&
$0.768\;(-0.49)$
&
$0.730\;(-0.33)$
&
$0.964\;(-0.63)$
&
$0.821\;(-0.30)$
\\
$1$
&
$0.893\;(0.17)$
&
$0.529\;(-0.16)$
&
$0.413\;(0.44)$
&
$0.362\;(-0.19)$
&
$0.206\;(0.47)$
\\
$10$
&
$0.988\;(0.94)$
&
$0.067\;(0.39)$
&
$0.136\;(0.28)$
&
$4.56\times10^{-3}\;(0.39)$
&
$1.87\times10^{-2}\;(0.28)$
\\
\hline\hline
\end{tabular}
\caption{
Characteristic minima of the reduced curvature purity, trace distance,
and displacement QFI for the three Fourier scales shown in
Fig.~\ref{fig:three_kappa_refined}. Numbers in parentheses denote the
e-fold $N$ at which the corresponding minimum occurs.
}
\label{tab:kappa_scale_minima}
\end{table*}
\begin{table*}[t]
\centering
\renewcommand{\arraystretch}{1.25}
\setlength{\tabcolsep}{5pt}
\begin{tabular}{c|c|cc|cc|cc|cc}
\hline\hline
$\kappa_t$
&
$\mathcal R_\mu$
&
$\mathcal R_{D,q}$
&
$\mathcal R_{D,p}$
&
$\mathcal R_{Q,q}$
&
$\mathcal R_{Q,p}$
&
$\mathcal B_q$
&
$\mathcal B_p$
&
$\mathcal Q_q$
&
$\mathcal Q_p$
\\
\hline
$0.1$
& $0.893$
& $0.664$
& $0.849$
& $0.193$
& $0.818$
& $0.029$
& $0.070$
& $0.045$
& $0.183$
\\

$1$
& $0.825$
& $0.622$
& $0.722$
& $0.488$
& $0.568$
& $0.174$
& $0.276$
& $0.356$
& $0.452$
\\

$10$
& $0.954$
& $0.834$
& $0.813$
& $0.605$
& $0.594$
& $0.608$
& $0.537$
& $0.605$
& $0.584$
\\
\hline\hline
\end{tabular}
\caption{ 
Recovery and revival measures for the three representative Fourier scales shown in Fig.~\ref{fig:three_kappa_refined}. The purity-recovery fraction $\mathcal R_\mu$, defined in Eq.~(\ref{eq:recoherence_fraction}), quantifies the fraction of the lost reduced curvature purity recovered by the final time $N_f$. The normalized quantities $\mathcal R_{D,j}$ and $\mathcal R_{Q,j}$, with $j\in\{q,p\}$, defined in Eq.~(\ref{eq:operational_recovery_fraction}), give  
the fractions of their respective maximum
losses recovered by the final time $N_f$.  The accumulated positive increments $\mathcal B_j$ and $\mathcal Q_j$, defined in Eq.~(\ref{eq:numerical_positive_increments}), quantify the total trace-distance and QFI revival, respectively, over all intervals of positive flow.
}
\label{tab:kappa_scale_recovery}
\end{table*}   
    
 For the super-Hubble case, $\kappa_t=0.1$, the reduced curvature state
remains close to pure, reaching
$\mu_{\mathcal S,k}^{\rm min}\simeq0.933$. Although the absolute purity
loss is modest, a large fraction of it is subsequently recovered,
$\mathcal R_\mu\simeq0.893$. The selected operational diagnostics are
also only weakly suppressed:
$D_q^{\rm min}\simeq0.768$,
$D_p^{\rm min}\simeq0.730$,
$F_{Q,k}^{(\delta_q),{\rm min}}\simeq0.964$, and
$F_{Q,k}^{(\delta_p),{\rm min}}\simeq0.821$.
The corresponding recovery fractions 
with respect to maximum suppression are
$(\mathcal R_{D,q},\mathcal R_{D,p})\simeq(0.664,0.849)$ and
$(\mathcal R_{Q,q},\mathcal R_{Q,p})\simeq(0.193,0.818)$, while the
accumulated revival measures remain small,
$(\mathcal B_q,\mathcal B_p)\simeq(0.029,0.070)$ and
$(\mathcal Q_q,\mathcal Q_p)\simeq(0.045,0.183)$.
The normalized recovery fractions associated with very small absolute
losses should, however, be interpreted with care. In particular, the
small value $\mathcal R_{Q,q}\simeq0.193$ results from the small initial
decrease of the $q$-oriented QFI and does not indicate a substantial
loss of displacement sensitivity. Overall, within the benchmark
evolution, a mode that is already super-Hubble at the peak of the turn
retains most of the distinguishability and displacement sensitivity
carried by the selected probes.\\

 For the near-Hubble case, $\kappa_t=1$, the reduced curvature purity
undergoes the largest suppression among the three representative
scales, reaching
$\mu_{\mathcal S,k}^{\rm min}\simeq0.893$, with a subsequent recovery
fraction $\mathcal R_\mu\simeq0.825$. The selected operational
diagnostics are also substantially suppressed:
$D_q^{\rm min}\simeq0.529$,
$D_p^{\rm min}\simeq0.413$,
$F_{Q,k}^{(\delta_q),{\rm min}}\simeq0.362$, and
$F_{Q,k}^{(\delta_p),{\rm min}}\simeq0.206$.
By the final time, approximately $62\%$ and $72\%$ of the lost
$q$- and $p$-oriented distinguishability are recovered,
$(\mathcal R_{D,q},\mathcal R_{D,p})\simeq(0.622,0.722)$,
while the corresponding QFI recovery fractions are
$(\mathcal R_{Q,q},\mathcal R_{Q,p})\simeq(0.488,0.568)$.
The accumulated positive increments,
$(\mathcal B_q,\mathcal B_p)\simeq(0.174,0.276)$ and
$(\mathcal Q_q,\mathcal Q_p)\simeq(0.356,0.452)$,
further indicate appreciable revival activity during the evolution.
Thus, within the three representative scales considered here, the
near-Hubble regime exhibits the strongest reduced-state mixedness,
accompanied by substantial and orientation-dependent loss and recovery
of the distinguishability and displacement sensitivity carried by the
selected probes.\\

For the sub-Hubble case, $\kappa_t=10$, the behavior is qualitatively
different. The reduced curvature state remains nearly pure,
$\mu_{\mathcal S,k}^{\rm min}\simeq0.988$, and recovers almost all of
this small purity loss, with $\mathcal R_\mu\simeq0.954$. In contrast,
the selected operational diagnostics undergo a pronounced suppression:
$D_q^{\rm min}\simeq0.067$ and
$D_p^{\rm min}\simeq0.136$, while
$F_{Q,k}^{(\delta_q),{\rm min}}\simeq4.6\times10^{-3}$ and
$F_{Q,k}^{(\delta_p),{\rm min}}\simeq1.9\times10^{-2}$.
A substantial fraction of these losses is subsequently recovered,
with
$(\mathcal R_{D,q},\mathcal R_{D,p})\simeq(0.834,0.813)$ and
$(\mathcal R_{Q,q},\mathcal R_{Q,p})\simeq(0.605,0.594)$.
The corresponding accumulated revival measures,
$(\mathcal B_q,\mathcal B_p)\simeq(0.608,0.537)$ and
$(\mathcal Q_q,\mathcal Q_p)\simeq(0.605,0.584)$,
are also the largest among the three representative scales.
Thus, within the benchmark evolution, a reduced curvature state can
remain almost pure while the distinguishability and displacement
sensitivity associated with the selected phase-space directions are
strongly suppressed and subsequently undergo substantial recovery.\\

The locations of the minima further show that the different diagnostics
are not temporally synchronized. In particular, for $\kappa_t=10$, the
trace distance and QFI reach their minima around
$N\simeq0.3$--$0.4$, whereas the purity minimum occurs later, near
$N\simeq0.94$. The RHP indicator provides a complementary map-level
perspective by identifying intervals of CP-indivisibility, which need
not coincide with extrema or revivals of the state- or probe-dependent
diagnostics. Taken together, the absolute losses, normalized recovery
fractions, and accumulated revival measures show that modes near Hubble
crossing are associated with stronger reduced-state mixedness, whereas
sub-Hubble modes can exhibit a much more pronounced transient
suppression and subsequent recovery of the distinguishability and
displacement sensitivity carried by the selected probes. Purity,
CP-divisibility, trace distance, and QFI therefore probe distinct but
complementary aspects of the reduced curvature dynamics.\\

\begin{figure}[!h]
\centering
\includegraphics[width=0.98\textwidth]{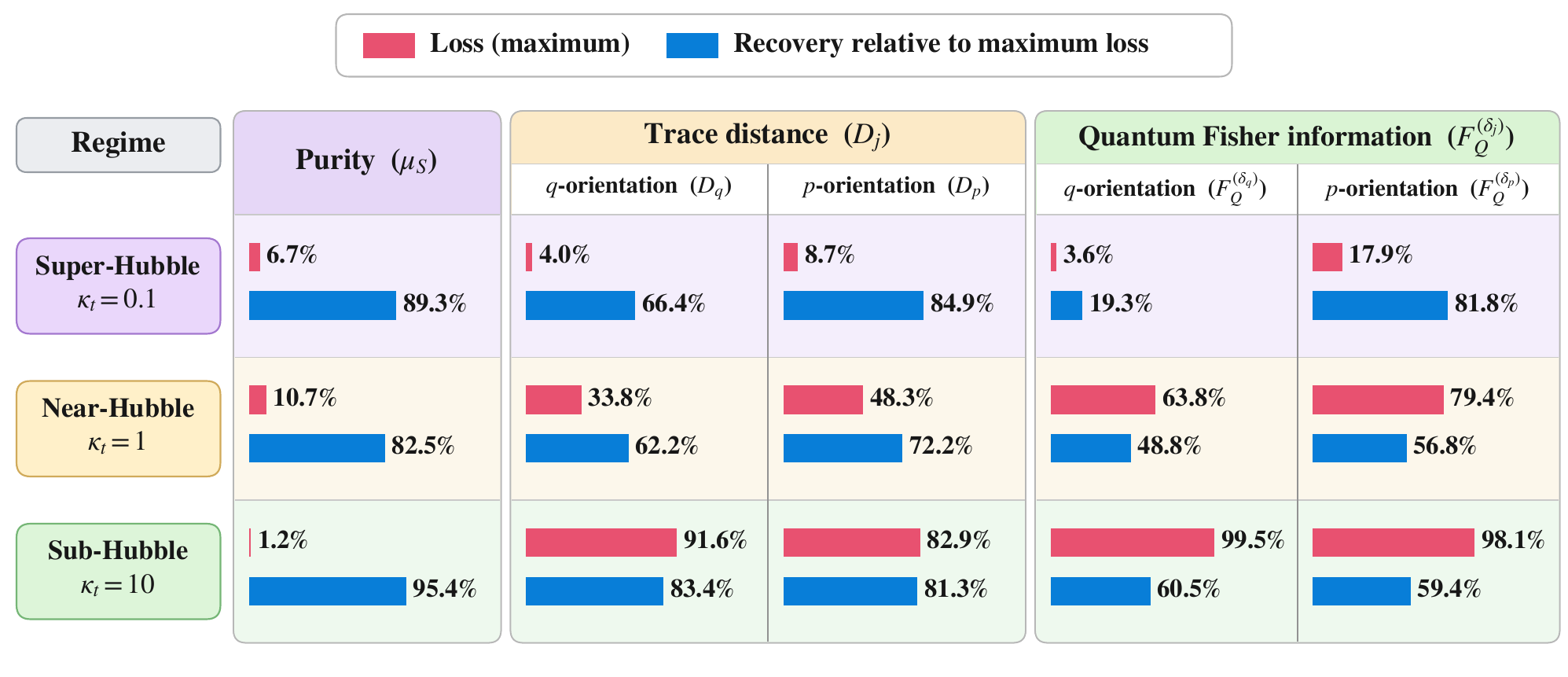}
\caption{
 Fourier-scale dependence of the maximum loss and subsequent
recovery of reduced curvature purity $\mu_S$, trace distances
$D_q$ and $D_p$, and displacement QFI $F_Q^{(\delta_q)}$ and
$F_Q^{(\delta_p)}$ for $\kappa_t=0.1$, $1$, and $10$.
The maximum loss (red) is expressed relative to the initial
value of each diagnostic, while the recovery (blue) denotes
the fraction of this loss recovered by the final epoch $N_f$.
All values are expressed as percentages.
}
\label{fig:intro_loss_recovery}
\end{figure}

\begin{figure}[!h]
\centering
\includegraphics[width=0.98\textwidth]{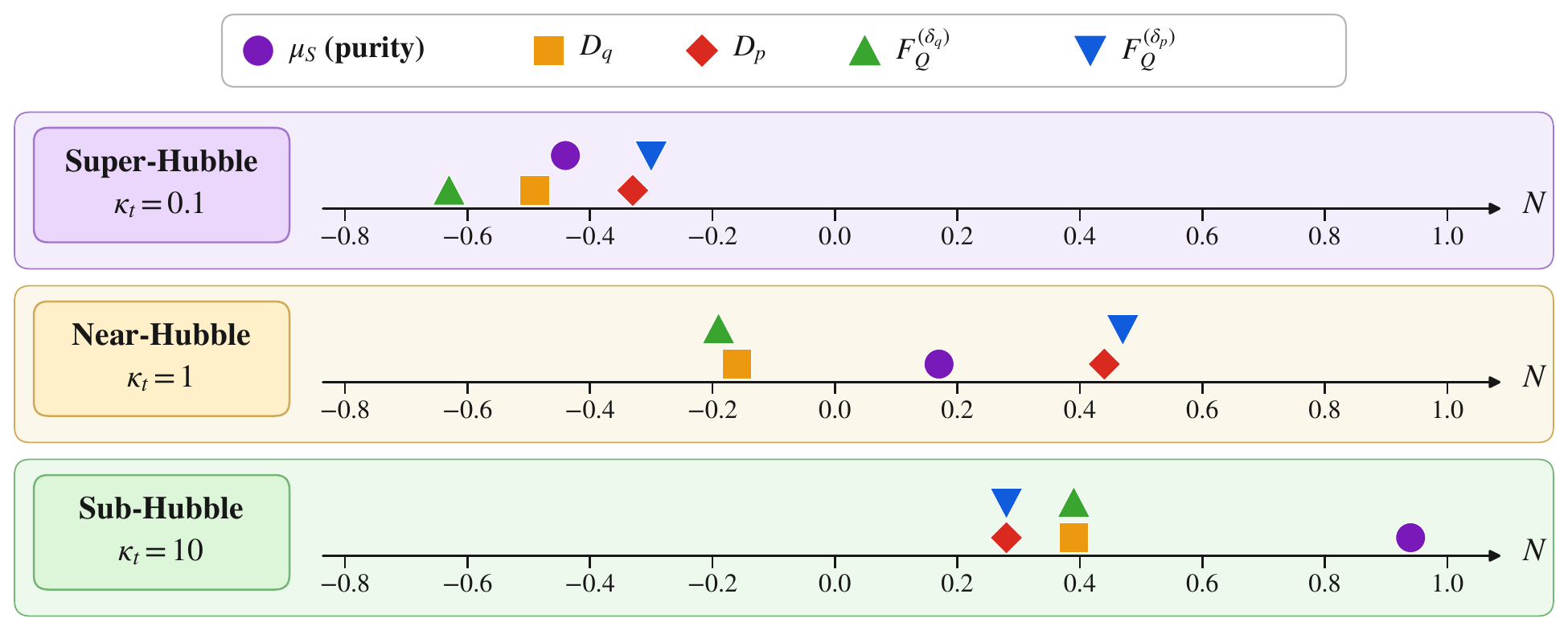}
\caption{
Relative epochs of maximal suppression of reduced curvature
purity, trace distance, and displacement QFI for the three
representative Fourier scales. Here, $N$ denotes the number
of e-folds relative to the epoch of maximum background
turning rate ($N=0$).
}
\label{fig:intro_epochs_minima}
\end{figure}

 A visual synthesis of these results is presented in
Figs.~\ref{fig:intro_loss_recovery} and
\ref{fig:intro_epochs_minima}, highlighting, respectively,
the magnitude of maximum loss and recovery relative 
to the maximum suppression and the relative epochs
of maximal suppression of the different diagnostics.
Together, these figures illustrate the inequivalence of
purity recovery and operational information recovery,
both in magnitude and in temporal occurrence, and reveal
their characteristic dependence on the Fourier scale and
the phase-space orientation of the initial encoding.

\subsection{Scale-resolved information dynamics in the $(N,\kappa_t)$ plane}
\begin{figure}[!h]
\centering

\begin{minipage}[t]{0.32\textwidth}
    \centering
    \includegraphics[width=\linewidth]
    {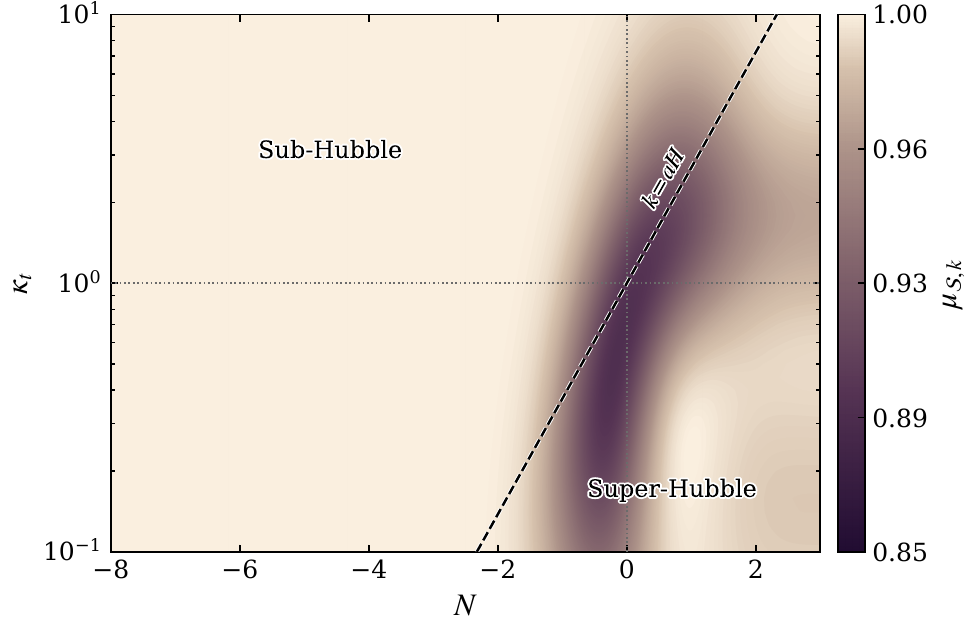}
    \par\vspace{0.15em}
    \small (a)
\end{minipage}
\hfill
\begin{minipage}[t]{0.32\textwidth}
    \centering
    \includegraphics[width=\linewidth]
    {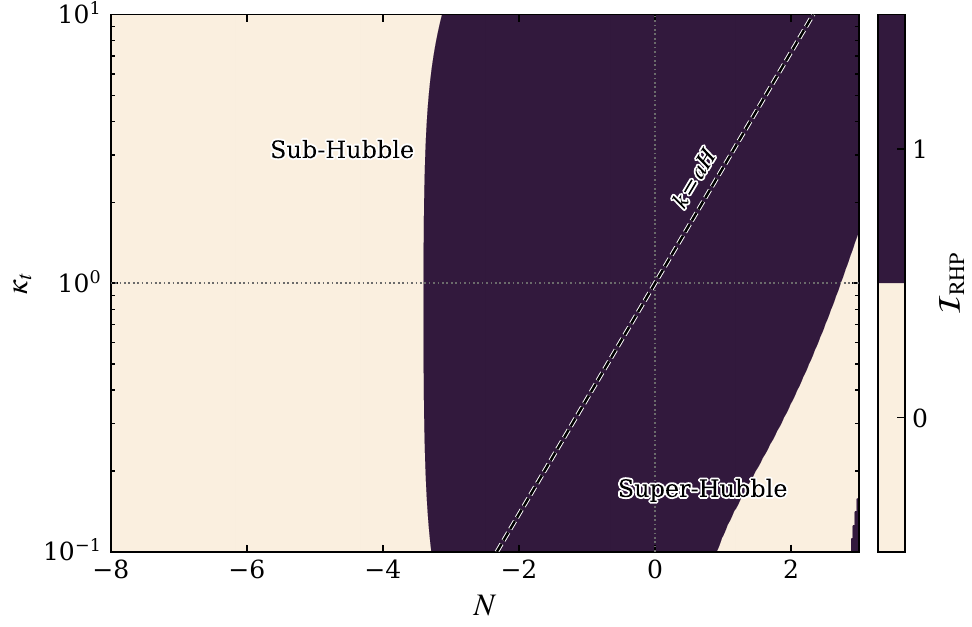}
    \par\vspace{0.15em}
    \small (b)
\end{minipage}
\hfill
\begin{minipage}[t]{0.32\textwidth}
    \centering
    \includegraphics[width=\linewidth]
    {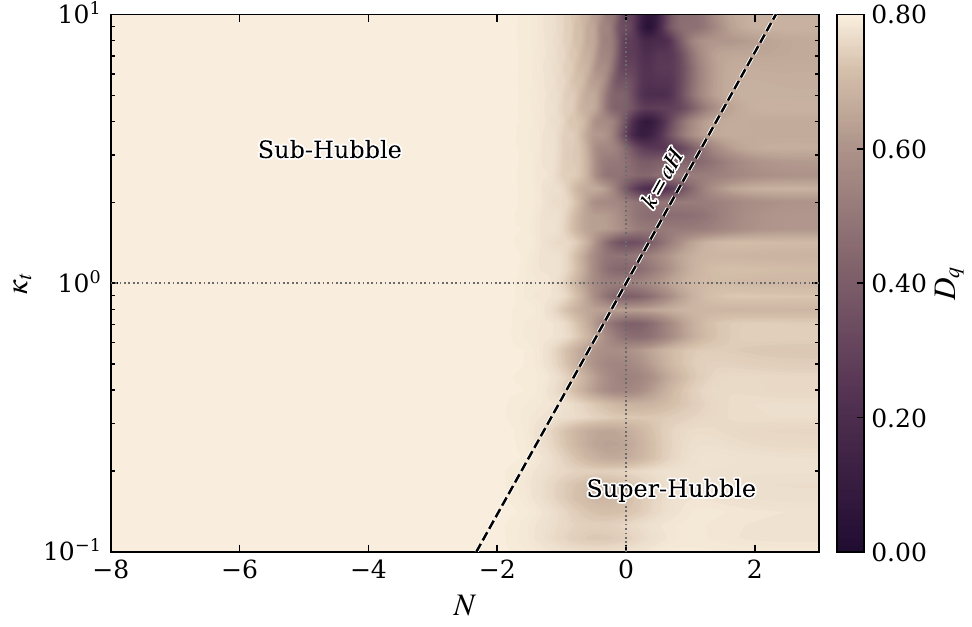}
    \par\vspace{0.15em}
    \small (c)
\end{minipage}

\par\vspace{0.6em}

\begin{minipage}[t]{0.32\textwidth}
    \centering
    \includegraphics[width=\linewidth]
    {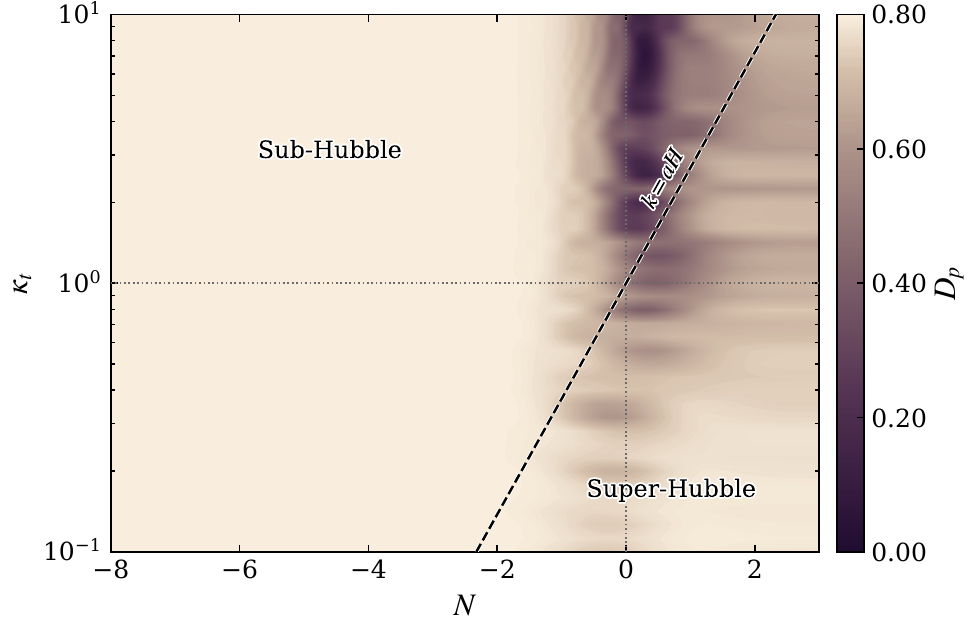}
    \par\vspace{0.15em}
    \small (d)
\end{minipage}
\hfill
\begin{minipage}[t]{0.32\textwidth}
    \centering
    \includegraphics[width=\linewidth]
    {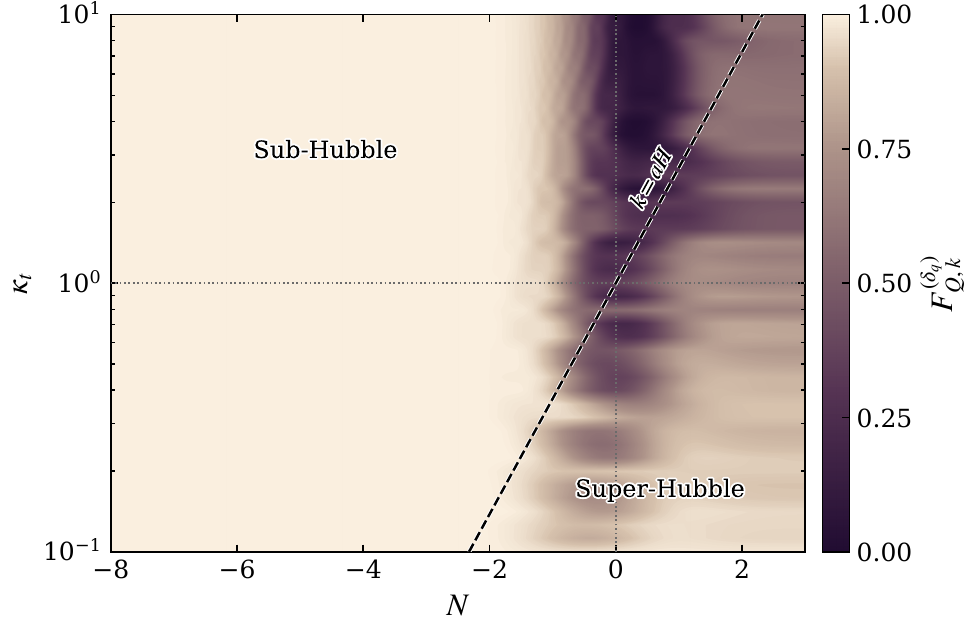}
    \par\vspace{0.15em}
    \small (e)
\end{minipage}
\hfill
\begin{minipage}[t]{0.32\textwidth}
    \centering
    \includegraphics[width=\linewidth]
    {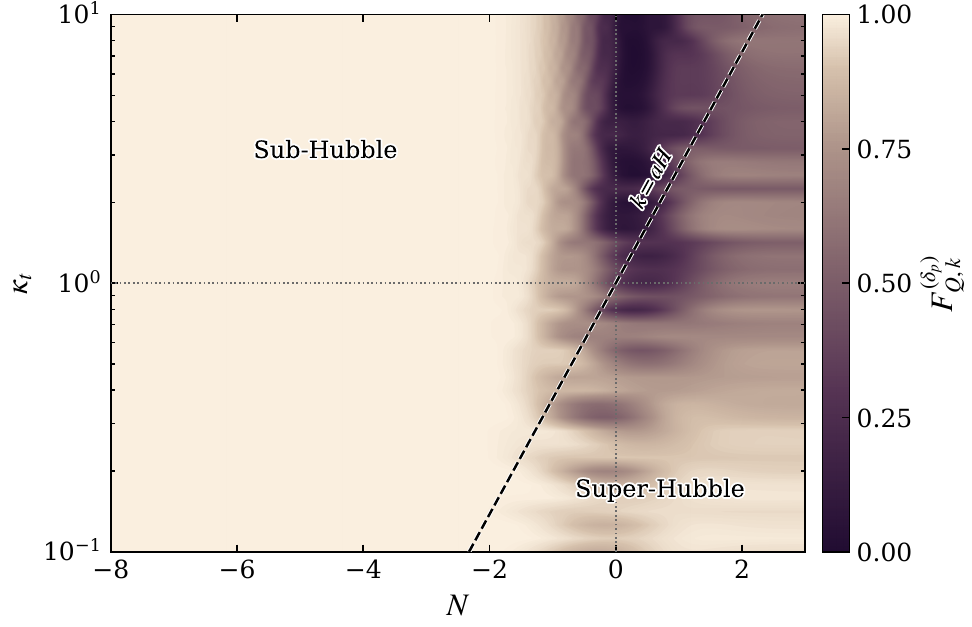}
    \par\vspace{0.15em}
    \small (f)
\end{minipage}

\caption{
Scale- and time-resolved information-theoretic diagnostics in the
$(N,\kappa_t)$ plane for the reference parameters
$\epsilon=0.01$, $M_{\mathcal F}/H=2$, $\alpha_0=1$, and
$\sigma_N=1$. Shown are (a) the reduced curvature purity
$\mu_{\mathcal S,k}$, (b) the binary finite-interval RHP
CP-indivisibility indicator $\mathcal I_{\rm RHP}$,
(c) and (d) the trace distances $D_q$ and $D_p$ for the
selected $q$- and $p$-oriented probe pairs, and
(e) and (f) the corresponding displacement QFI
$F_{Q,k}^{(\delta_q)}$ and $F_{Q,k}^{(\delta_p)}$.
The diagonal dashed line, $\kappa_t=e^{(1-\epsilon)N}$,
represents the Hubble-crossing condition $k=aH$ (see Eq.\ \eqref{eq:kappa_evolution}),
separating the sub-Hubble ($k>aH$) and super-Hubble
($k<aH$) regimes above and below the boundary,
respectively. The vertical dotted line denotes the
peak of the turn at $N=0$, while the horizontal dotted
line marks $\kappa_t=1$, corresponding to Hubble crossing
at the peak of the turn. 
}
\label{fig:N_kappa_contours}
\end{figure}

To extend the analysis beyond the three representative modes, we examine the diagnostics over the continuous $(N,\kappa_t)$ plane, varying $0.1\leqslant\kappa_t\leqslant10$ while keeping all other parameters fixed at the reference values of Eq.~(\ref{eq:benchmark_parameters}). Figure~\ref{fig:N_kappa_contours} thus provides a scale-resolved view of the reduced dynamics and highlights the distinct regions in which the different information-theoretic signatures become prominent. 
The purity loss is concentrated near the turn for $\kappa_t=\mathcal O(1)$, with the minimum occurring at $\kappa_t\simeq0.56$ and $N\simeq-0.13$. Thus, the enhanced mixedness seen at $\kappa_t=1$ is part of a broader Hubble-scale feature rather than an isolated mode-specific effect. 
The trace-distance and QFI suppression is strongest in the sub-Hubble regime, occurring mainly around and shortly after the turn. Within the scan, the deepest $q$-oriented suppression appears near $\kappa_t\simeq10$ and $N\simeq0.39$, while the corresponding $p$-oriented minimum occurs near $\kappa_t\simeq8.3$ and $N\simeq0.30$. The offset between the two orientations reflects the directional dependence of the operational information encoded in phase space.
At larger $\kappa_t$, oscillatory sub-Hubble dynamics produces pronounced scale-dependent modulations in the operational diagnostics. Notably, the curvature mode remains nearly pure over much of this region while the trace distance and displacement QFI are strongly suppressed, demonstrating that reduced-state mixedness alone does not determine the operational accessibility of information encoded in a given phase-space direction. 
The RHP indicator exhibits a broader pattern of CP-indivisibility that does not generally coincide with the strongest purity, trace-distance, or QFI signatures, consistent with its characterization of the dynamical map rather than a particular state or probe. Overall, the continuous scan shows that reduced-state mixedness is most pronounced near Hubble-scale modes, whereas the strongest transient loss and recovery of operational information occurs deeper in the sub-Hubble regime, confirming that the different diagnostics probe complementary aspects of the curvature--entropic dynamics. 

 \begin{figure}[!h]
\centering

\begin{minipage}[t]{0.32\textwidth}
    \centering
    \includegraphics[width=\linewidth]
    {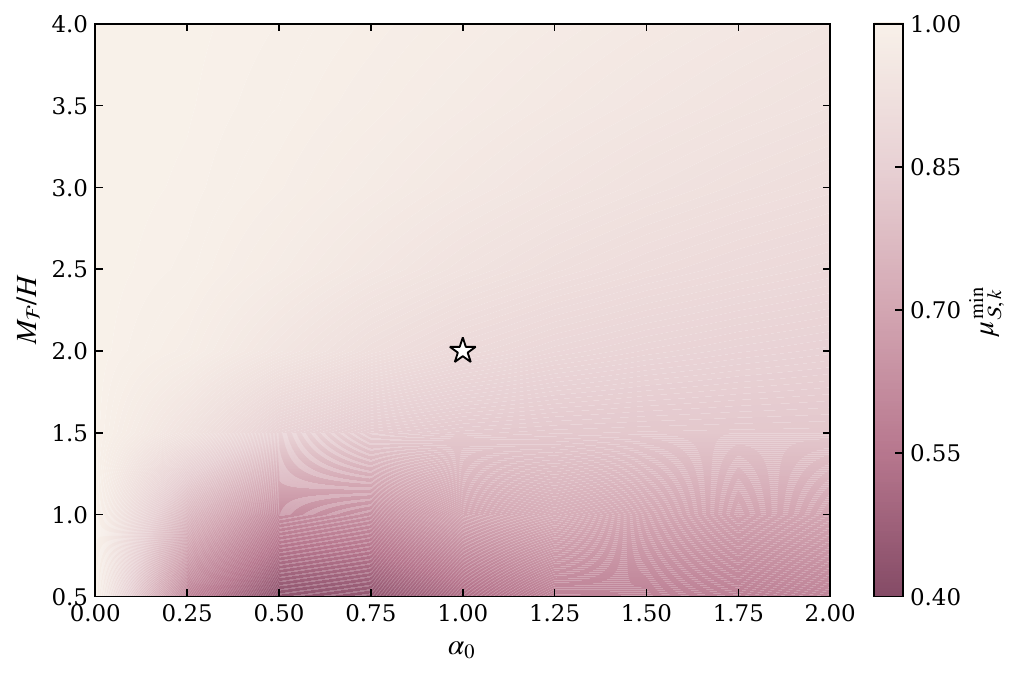}
    \par\vspace{0.15em}
    \small (a)
\end{minipage}
\hfill
\begin{minipage}[t]{0.32\textwidth}
    \centering
    \includegraphics[width=\linewidth]
    {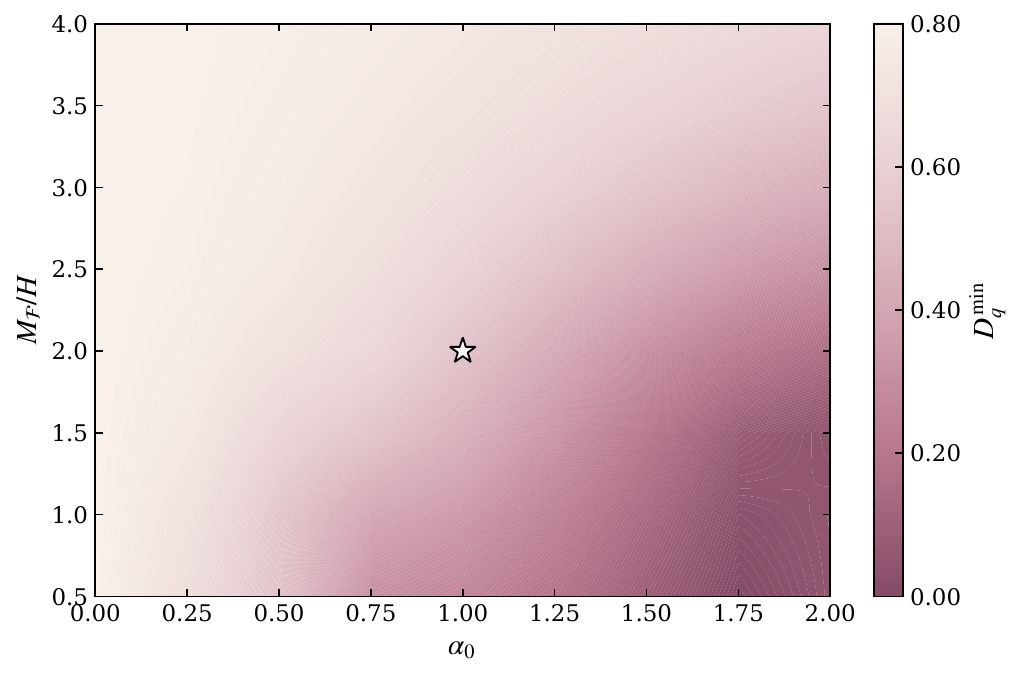}
    \par\vspace{0.15em}
    \small (b)
\end{minipage}
\hfill
\begin{minipage}[t]{0.32\textwidth}
    \centering
    \includegraphics[width=\linewidth]
    {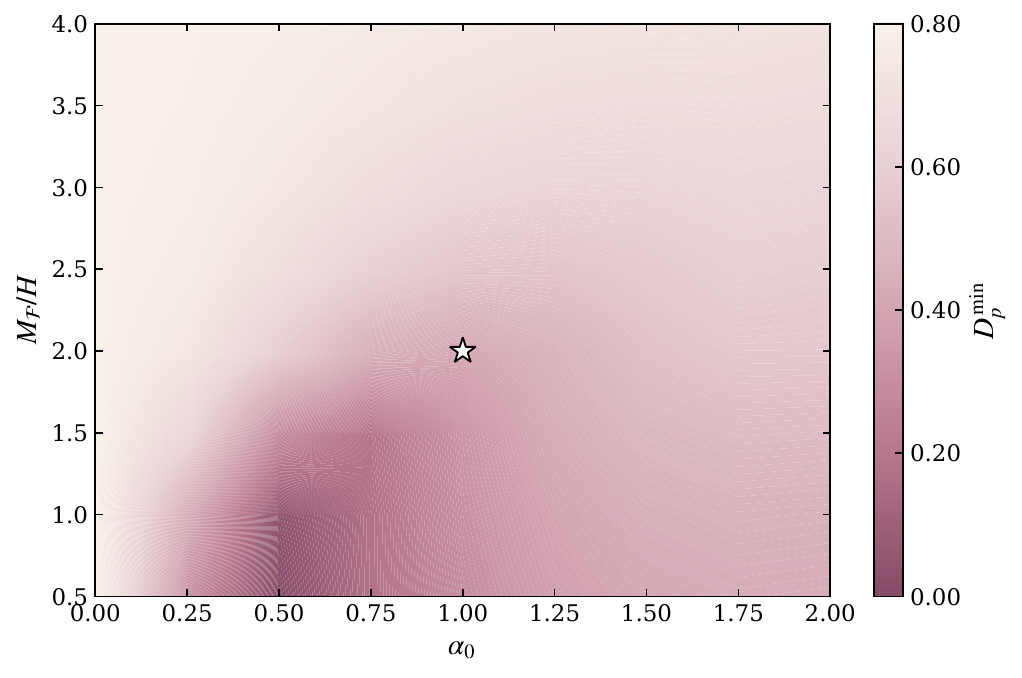}
    \par\vspace{0.15em}
    \small (c)
\end{minipage}

\par\vspace{0.6em}

\begin{minipage}[t]{0.32\textwidth}
    \centering
    \includegraphics[width=\linewidth]
    {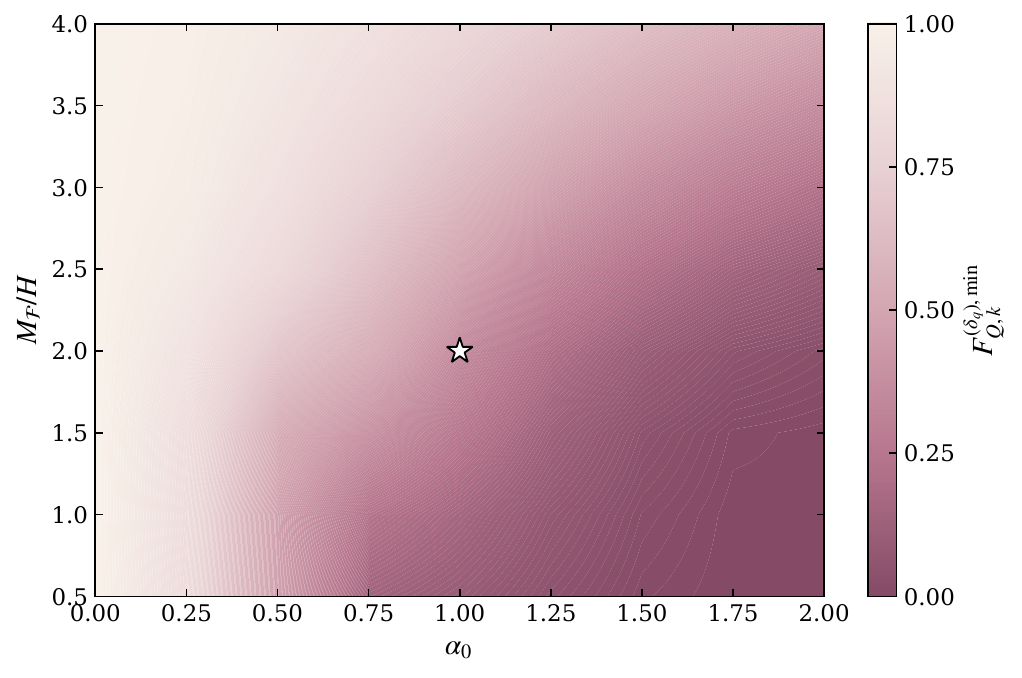}
    \par\vspace{0.15em}
    \small (d)
\end{minipage}
\hspace{0.02\textwidth}
\begin{minipage}[t]{0.32\textwidth}
    \centering
    \includegraphics[width=\linewidth]
    {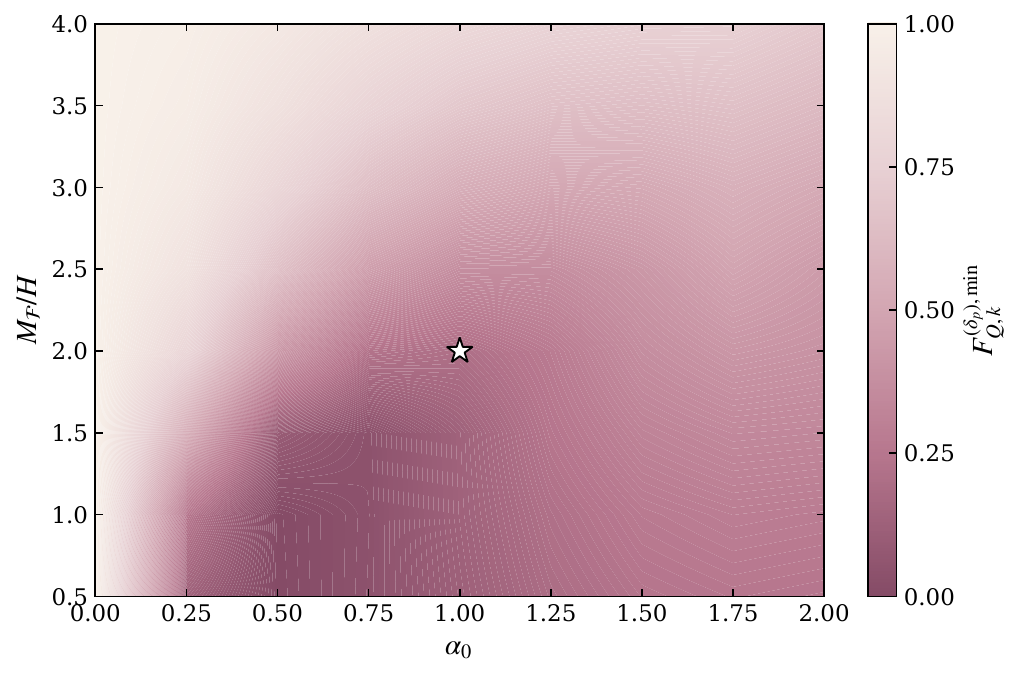}
    \par\vspace{0.15em}
    \small (e)
\end{minipage}

\caption{Dependence of the reduced curvature dynamics on the turn amplitude
$\alpha_0$ and entropic mass ratio $M_{\mathcal F}/H$ at fixed $\kappa_t=1$,
$\epsilon=0.01$, and $\sigma_N=1$. At each point in the parameter plane,
the colour gives the minimum sampled over $N\in[-8,3]$ of the quantity
shown in that panel: (a) the reduced curvature purity $\mu_{\mathcal S,k}$,
(b) and (c) the trace distances $D_q$ and $D_p$, and (d) and (e) the
displacement QFI $F_{Q,k}^{(\delta_q)}$ and $F_{Q,k}^{(\delta_p)}$,
respectively. The star in each panel marks the benchmark
$(\alpha_0,M_{\mathcal F}/H)=(1,2)$. The colour values at the stars in
panels (a)--(e) correspond, in order, to the minima marked by grey dots
on the orange $\kappa_t=1$ curves in Fig.\ref{fig:three_kappa_refined}.}
\label{fig:alpha_mass_plane}
\end{figure}

\subsection{Dependence on entropic mass and turn strength}
To complement the Fourier-scale analysis, we fix $\kappa_t=1$
and explore the dependence on the turn amplitude $\alpha_0$
and entropic mass $M_{\mathcal F}/H$.
Figure~\ref{fig:alpha_mass_plane} maps the minima of the
reduced purity, trace distance, and displacement QFI over
the evolution interval, identifying where curvature--entropic
coupling produces the strongest loss of reduced-state purity,
distinguishability, and displacement sensitivity, respectively. 
The largest purity reduction occurs for a relatively light
entropic mode, with $\mu_{\mathcal S,k}^{\rm min}\simeq0.806$
near $(\alpha_0,M_{\mathcal F}/H)=(0.9,1.4)$.
Remarkably, the operational diagnostics exhibit a different,
orientation-dependent response: at $M_{\mathcal F}/H=1.4$,
the $q$-oriented trace distance and QFI attain minima
$(D_q^{\rm min},F_{Q,k}^{(\delta_q),{\rm min}})
\simeq(0.189,0.043)$ near $\alpha_0=1.5$,
whereas their $p$-oriented counterparts reach
$(D_p^{\rm min},F_{Q,k}^{(\delta_p),{\rm min}})
\simeq(0.089,0.0094)$ near $\alpha_0=0.6$.
The separation of these extrema reveals that the strongest
generation of reduced-state mixedness and the greatest loss
of operational information need not occur in the same
parameter regime, highlighting the phase-space anisotropy
arising from the interplay of inflationary squeezing and
curvature--entropic coupling.

\subsection{Numerical consistency and interpretation}
The parameter-space structures remain stable under refinement of the propagation accuracy and parameter-grid resolution. As a direct check, the numerical evolution was compared with the adaptive eighth-order DOP853 integration of Eq.~(\ref{eq:numerical_propagator}), yielding
$\max|\Delta\mu_{\mathcal S,k}|\lesssim4\times10^{-7}$ and
$\max|\Delta D_j|,\max|\Delta F_{Q,k}^{(\delta_j)}|\lesssim5\times10^{-5}$
for the tested trajectories, with no change in the finite-interval RHP classification. Preservation of the canonical symplectic structure was monitored using Eq.~(\ref{eq:symplectic_error}), with
$\epsilon_{\rm symp}^{\rm max} \leqslant 10^{-10}$.
The Robertson--Schr\"odinger physicality condition and the reduced-channel complete-positivity condition in Eq.~(\ref{eq:gaussian_channel_CP}) were satisfied throughout the scan up to numerical residuals of order $10^{-12}$. Convergence of the trace-distance reconstruction in Eq.~(\ref{eq:fock_trace_distance}) was checked by increasing $n_{\rm cut}$; values $n_{\rm cut}\leqslant34$ were sufficient over the full scan, with discarded occupation weight below $7\times10^{-16}$ and changes in $D_k$ below the adopted $10^{-8}$ tolerance. The displacement QFI obtained from Eq.~(\ref{eq:displacement_qfi}) was additionally checked against the fidelity relation in Eq.~(\ref{eq:fid1}) and the Fuchs--van de Graaf bounds in Eq.~(\ref{eq:fuchs_van_de_graaf}).\\

These checks indicate that the structures observed in the $(\alpha_0,M_{\mathcal F}/H)$ plane are robust against the numerical choices used in the calculation. For the selected displaced probes, the trace distance $D_j$ and the displacement QFI $F_{Q,k}^{(\delta_j)}$ display closely related behavior, consistent with the exact relation between QFI and fidelity in Eq.~(\ref{eq:fid1}) and the bounds in Eq.~(\ref{eq:fuchs_van_de_graaf}). Their extrema nevertheless need not coincide with the minimum of the reduced purity in Eq.~(\ref{eq:gaussian_purity}), since purity characterizes the overall mixedness of the curvature mode rather than the information encoded in a particular probe direction. The RHP criterion is different in character: through the intermediate-map condition in Eq.~(\ref{eq:intermediate_cp}) and the indicator in Eq.~(\ref{eq:rhp_binary_indicator}), it probes CP-divisibility of the reduced dynamical map rather than a particular state or encoding. Purity, trace distance, QFI, and the RHP indicator therefore provide complementary characterizations of the reduced curvature dynamics.

\section{Conclusions and outlook}
\label{sec:conclusion}

In this work, we have investigated the relationship between
cosmological recoherence, information recovery, and quantum
non-Markovianity in the reduced dynamics of primordial
curvature perturbations during two-field inflation.
Exploiting the symplectic structure of the coupled
curvature--entropic dynamics, we have obtained an exact
Gaussian-channel description of the reduced curvature
evolution within the quadratic theory, without invoking
weak-coupling or Markovian approximations.
A distinctive aspect of our approach is the systematic
comparison of reduced-state purity, CP-divisibility,
state distinguishability, and metrological sensitivity
within a common dynamical framework.
In particular, the use of covariance-normalized displaced
Gaussian states places the trace-distance and QFI diagnostics
on an equal footing, enabling a direct comparison of
their dynamical behavior for the same family of initial
state preparations.\\

Our results reveal a pronounced dependence of the
information-theoretic properties of the reduced curvature
dynamics on the physical wave number of the perturbation
relative to the Hubble scale at the epoch of
curvature--entropic interaction.
For the benchmark inflationary background, modes that
are super-Hubble at the peak of the turn largely retain
their initially encoded information, whereas those near
Hubble crossing exhibit stronger curvature--entropic
entanglement.
A qualitatively distinct behavior emerges in the
sub-Hubble regime, where an almost complete transient
suppression of displacement sensitivity and a pronounced
loss of preparation distinguishability occur despite
the reduced curvature state remaining nearly pure.
For the representative mode $\kappa_t=10$, the maximum
purity loss is approximately $1.2\%$, whereas the
displacement QFI is suppressed by more than $98\%$
along both encoding directions.
Moreover, the nearly complete recovery of the lost
purity is accompanied by only partial recovery of
distinguishability and displacement sensitivity.
These results demonstrate that neither the preservation
nor the restoration of reduced-state purity necessarily
implies the corresponding preservation or recovery of
initially encoded information.  The numerical analysis further reveals that the magnitude of transient information loss does not necessarily determine the extent of its subsequent recovery. For instance, the $q$-encoded QFI recovers only $19.3\%$ of its loss at $\kappa_t=0.1$, despite a relatively shallow minimum of $0.964$, whereas a substantially deeper minimum of $4.56\times10^{-3}$ at $\kappa_t=10$ is followed by a recovery of $60.5\%$. This illustrates the absence of a straightforward correspondence between information loss and recovery across different Fourier scales. \\

The distinction between these diagnostics is further
manifested in their temporal behavior and dependence on
the interaction parameters.
The epochs of maximal suppression depend on the Fourier
scale and, for the operational diagnostics, on the
phase-space orientation of the initial encoding.
In particular, for the representative sub-Hubble mode,
the trace distance and displacement QFI attain their
minima considerably earlier than the reduced purity.
The extended Fourier-scale analysis confirms that the
separation between enhanced reduced-state mixedness near
the Hubble scale and pronounced operational information
loss in the sub-Hubble regime persists beyond the three
representative modes.
Furthermore, the joint variation of the entropic mass
and turn amplitude reveals distinct parameter
dependencies for the different diagnostics, with the
strongest purity loss and the greatest suppression of
distinguishability and displacement sensitivity
occurring in different regions of parameter space.
These results demonstrate that the conditions favoring
curvature--entropic entanglement generation need not
coincide with those favoring the loss of operationally
accessible information.\\

For the selected Gaussian displacement probes, we have
established an exact relation between the displacement QFI
and the squared Uhlmann fidelity, thereby connecting
metrological sensitivity to preparation distinguishability.
Once the reduced probes become mixed, however, their
trace distance is no longer uniquely determined by the
displacement QFI, rendering the two diagnostics
complementary rather than interchangeable.
Furthermore, although revivals of trace distance and QFI
provide operational witnesses of CP-indivisibility,
the latter need not be accompanied by revivals for
the selected probes.
The map-level characterization of quantum
non-Markovianity thus complements the state- and
preparation-dependent descriptions of the reduced
curvature dynamics.
These signatures should, however, be distinguished from
genuine causal information backflow from the entropic
sector, whose identification requires an analysis of
the underlying multitime quantum dynamics.\\

Taken together, our findings demonstrate that cosmological
recoherence and operational information recovery constitute
distinct aspects of the reduced quantum dynamics of
primordial perturbations, with their interplay depending
sensitively on the Fourier scale relative to the Hubble
scale at the epoch of interaction.
The combined information-theoretic framework developed
here reveals features of curvature--entropic dynamics
that cannot be inferred from purity dynamics alone.
In particular, neither the preservation nor the
restoration of reduced-state purity is sufficient to
infer the preservation or recovery of initially encoded
information.
This distinction underscores the importance of
complementing purity-based characterizations of
cosmological decoherence and recoherence with operational
diagnostics of quantum information.\\

The present analysis is restricted to the dynamics of
cosmological perturbations at quadratic order on a
phenomenologically prescribed inflationary background.
Natural extensions of the present framework include
background trajectories derived from specific multifield
inflationary potentials, additional entropic degrees of
freedom, and nonlinear interactions between perturbation
modes.
Another important direction is to investigate the causal
structure underlying information recovery through a
multitime characterization of the reduced curvature
dynamics.
Of particular interest is whether the scale-dependent
information-theoretic signatures identified here can be
related to observable primordial correlation functions.
These extensions may help elucidate the role of quantum
information loss and recovery during inflation and their
implications for the quantum-to-classical transition of
primordial perturbations.

\vspace{1em}

\paragraph*{Acknowledgement} Authors would like to thank Anumita Mukhopadhyay for helpful discussions on the manuscript.
 \vspace{1em}

\paragraph*{Data Availability Statement} No external datasets were used in this study. The numerical results presented in the figures were generated from the model and equations described in the manuscript.
\vspace{1em}
\paragraph*{Use of Artificial Intelligence}
The authors used ChatGPT (OpenAI) to assist with language editing, \LaTeX{} formatting, and draw some figures. The authors take responsibility for the analysis, interpretation, and final text.

\bibliographystyle{JHEP} 
\bibliography{references}
\end{document}